\RequirePackage{lineno}
\documentclass[aps,prd,twocolumn,showpacs,superscriptaddress,letterpaper,asmmath,amssymb]{revtex4}

\usepackage{graphicx}
\usepackage{dcolumn}
\usepackage{bm}
\usepackage{feynmf}

\RequirePackage{xspace}
\usepackage{relsize}

\def\Kbar  {\kern 0.2em\overline{\kern -0.2em K}{}\xspace}

\def\Bbar    {\kern 0.18em\overline{\kern -0.18em B}{}\xspace}

\def\Qbar    {\kern 0.08em\overline{\kern -0.08em Q}{}\xspace}

\newcommand{\mev}{\ensuremath{\mathrm{\,Me\kern -0.1em V}}\xspace}
\newcommand{\mevc}{\ensuremath{{\mathrm{\,Me\kern -0.1em V\!/}c}}\xspace}
\newcommand{\mevcc}{\ensuremath{{\mathrm{\,Me\kern -0.1em V\!/}c^2}}\xspace}
\newcommand{\gev}{\ensuremath{\mathrm{\,Ge\kern -0.1em V}}\xspace}
\newcommand{\gevc}{\ensuremath{{\mathrm{\,Ge\kern -0.1em V\!/}c}}\xspace}
\newcommand{\gevcnospace}{\ensuremath{{\mathrm{\,Ge\kern -0.1em V\!/}c}}}
\newcommand{\gevcc}{\ensuremath{{\mathrm{\,Ge\kern -0.1em V\!/}c^2}}\xspace}
\newcommand{\bea}{\begin{eqnarray}}
\newcommand{\eea}{\end{eqnarray}}

\begin{document}
\setlength{\unitlength}{1mm}
\bibliographystyle{apsrev}

\title{Searching For Resonances inside Top-like Events}

\author{Jared~A.~Evans}
\affiliation{NHETC, Rutgers University, Piscataway, New Jersey}
\author{Ben~Kilminster}
\affiliation{Fermi National Accelerator Laboratory, Batavia, Illinois}
\author{Markus~A.~Luty}
\affiliation{University of California, Davis, Davis, California}
\author{Daniel~Whiteson}
\affiliation{University of California, Irvine, Irvine, California}


\begin{abstract}
In extended Higgs sectors, 
heavy Higgs bosons can decay via cascades
 to a light Higgs boson plus $W$ and $Z$ bosons.
We study signals of such sectors at the Tevatron and LHC that result
from resonant production of a heavy $H^0$ followed by the decay
$H^0 \to H^\pm W^\mp$
with 
$H^+ \to W^+ h^0 \to W^+ b\bar{b}$ or
$H^+ \to t\bar{b} \to W^+ b\bar{b}$.
The final states have the same particle content as that of $t\bar{t}$ production,
but with a resonant structure that can be used to distinguish signal events
from background events.
We propose analysis techniques and estimate the experimental
sensitivity of the Tevatron and LHC experiments to these signals.
\end{abstract}



\maketitle

\date{\today}

\section{Introduction}

The experimental study of the electroweak symmetry breaking sector
(or Higgs sector)
is one of the main goals of the experimental high energy physics program.
The minimal standard model with a single scalar Higgs boson is compatible
with all existing data if the Higgs boson mass is in the range 
115--140~GeV, and in fact intriguing hints of a Higgs boson
with mass near 125~GeV have recently been found at the CERN Large Hadron
Collider (LHC) \cite{ATLASHiggshint,CMSHiggshint}.
If a light Higgs-like state is definitively discovered, the next step
in the experimental program will
be to determine whether this state is in fact the Higgs
boson of the minimal standard model, part of an extended Higgs
sector (such as that of the minimal supersymmetric standard model, or
MSSM \cite{mssm}),
a composite Higgs \cite{comphiggs},
or a completely different
particle with Higgs-like couplings
(such as a radion in warped extra dimensions \cite{radionHiggs}
or dilaton \cite{dilatonHiggs}).

In this paper, we consider models with a light neutral Higgs
boson that is part of an extended Higgs sector.
Rather than assume a particular theoretical framework (such as the
MSSM), we take a phenomenological approach, using a general
2-Higgs doublet model as a convenient simplified
model \cite{simpmodels} to parameterize the signals.
This approach motivates a variety of signals with final states
involving the heaviest standard model particles
($W$, $Z$, $t$, and $b$), which have the strongest couplings
to the Higgs sector \cite{spin0res,spinWZtbsimpmodel}.
The $WW$ final state is enhanced by $WW$ scattering
in models where the Higgs sector
is strongly coupled \cite{strongww},
and this signal has been the subject of much detailed investigation
\cite{wwscatteringLHC}.
The phenomenology of resonant production of the
final states $Zh^0$ \cite{ZH} and $W^+ W^- Z$ \cite{WWZ}
have also been investigated.

In this paper, we focus on the final state $W^+ W^- b\bar{b}$,
which can have a large production rate from the process
$gg \to H^0$ 
followed by $H^0 \to H^\pm W^\mp$
with 
$H^+ \to W^+ h^0 \to W^+ b\bar{b}$ or
$H^+ \to t\bar{b} \to W^+ b\bar{b}$.
The main challenge with these signals is the
large $t\bar{t}$ background, which shares the same final state.
The Tevatron has performed in-depth studies of the properties of the top
quark  in the $t\bar{t}$ mode, including searches for charged Higgs bosons
in top-quark decay \cite{chiggsintop}, a measurement of the polarization
of the top-quark decay products \cite{toppol},
and independent measurements of the top-quark mass in different
top-quark decay modes~\cite{topmass}.
The model considered in this paper can be viewed as a continuation
of these investigations, understanding to what degree the $t\bar{t}$
sample at the Tevatron and the LHC
can contain final states with a $b\bar{b}$ resonance.

This paper is organized as follows.
In section II of the paper, we discuss a simplified model 
and Monte Carlo generation of the signal.
In section III, we address selection criteria and the dominant backgrounds.  Section IV and V discuss search strategies and expected sensitivity in $b\bar{b}$ and $Wb\bar{b}$ resonances respectively.

\section{Simplified Model}

To carry out this study, we use a simplified model that contains
only the minimal particle content necessary to describe the signal of
interest.
It contains two
neutral Higgs bosons, heavy ($H^0$) and light ($h^0$), and two 
charged Higgs particles ($H^{\pm}$).  
The parameters of the model are simply the masses of the Higgses 
and the production cross section times branching ratios for the 
assumed decays.
This allows for the simple interpretation of the parameters and the later mapping
to the specific parameters of more complete theories.

The production mode studied here is gluon-gluon fusion $gg \to H^0$.
We consider only the decay $H^0 \to H^\pm W^\mp$ and $h^0 \to b\bar{b}$.
We consider two possible decays of the charge Higgs state,
$H^+ \to W^+ h^0$ and $H^+ \to t\bar{b}$, performing two separate
analyses for the limiting cases where one of these dominates.
Although we are not committing to a specific model, it is worth
noting that in 2-Higgs doublet models, these decays are controlled
by the couplings
\begin{equation}
g_{H^+ W^- h^0} = \frac{2m_Z \cos(\beta - \alpha)}{v \cos \theta_W} ,
\ \ 
g_{H^+ \bar{t} b} = \frac{m_t}{v \tan\beta},
\end{equation}
which depend on different combinations of the angles
$\alpha$ and $\beta$ (defined for example in Ref.~\cite{HiggsHuntersGuide}).
These searches therefore probe different regions of parameter space
of this model.
On the other hand, the production cross section depends on the
coupling
\begin{equation}
g_{H^0 t\bar{t}} = \frac{m_t \sin\alpha}{v \sin\beta}.
\end{equation}
For parameters where this is unsuppressed,
the $H^+ \to t\bar{b}$ decay is also
unsuppressed, so in a 2-Higgs doublet model the modes
$H^+ \to W^+ h^0$ and $H^+ \to t\bar{b}$ are always both allowed.
Nevertheless, we will study these modes independently in this preliminary
study.

\begin{figure}[h]
\begin{center}
\includegraphics[width=0.8\linewidth]{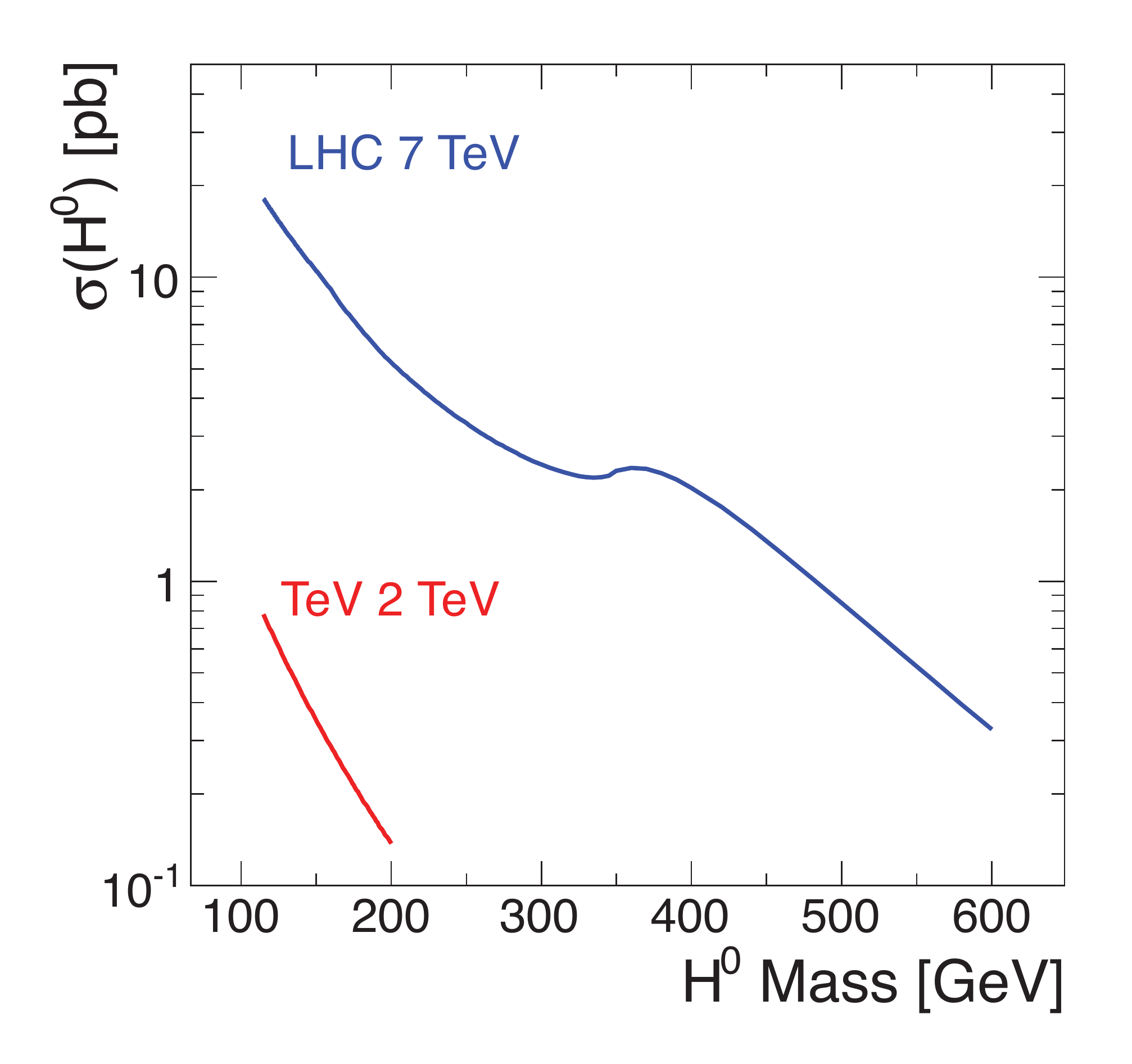}
\end{center}
\caption{Production cross section at NLO for $H^0$ as a function of mass
in a 2-Higgs doublet model with $\sin\alpha/\sin\beta = 1$, equivalent
to the SM Higgs production cross section,
for Tevatron $p\bar{p}$ collisions at $\sqrt{s}=1.96$~TeV~\cite{xsectev}, and for
LHC $pp$ collisions at $\sqrt{s}=7$~TeV~\cite{xseclhc}.}
\label{fig:xsec}
\end{figure}

To avoid unnecessary model-dependent assumptions in our results, we give 
expected limits on the $H^0$
production cross section times the product of the
appropriate branching ratios.
To interpret the results, we need to know the expected rates
in some reasonable models, so we give the cross section for
$H^0$ production  in Fig.~\ref{fig:xsec}, which is equivalent to the
SM Higgs production cross section if $\sin\alpha/\sin\beta = 1$.

Both signal and background events are generated with 
{\sc Madgraph}~\cite{madgraph}, while top-quark and $W$ boson decay,
showering and hadronization is performed by {\sc Pythia}~\cite{pythia}.
We use {\sc pgs}~\cite{pgs} tuned for Tevatron or ATLAS to provide
detector simulation.

\section{Selection and Backgrounds}

The event selection is similar to the standard single lepton 
selection used for Tevatron and LHC measurements \cite{ljets1,ljets2} 
of the $t\bar{t} \rightarrow W^{+}\bar{b} W^{-}b$ final states, with one $W$ 
boson decaying leptonically.  We require:
\begin{itemize}
\item exactly one electron or muon, with $p_T>20$ GeV and $|\eta|<2.5$
\item at least four jets, each with $p_T>20$ GeV and $|\eta|<2.5$
\item at least 20~GeV of missing transverse momentum
\item at least one $b$-tagged jet
\end{itemize}

The $b$-tagging algorithm parametrized in PGS is 36\% (44\%) efficient per $b$-quark jet 
for the Tevatron (LHC). The dominant standard model background is $t\bar{t}$ production.
At the Tevatron (LHC) $W+$jets contributes 25\% (10\%) after this selection.
In this study, we consider only the $t\bar{t}$ background.

\section{Resonances in $b\bar{b}$}

\begin{figure}[h]
\begin{center}
\includegraphics[width=0.8\linewidth]{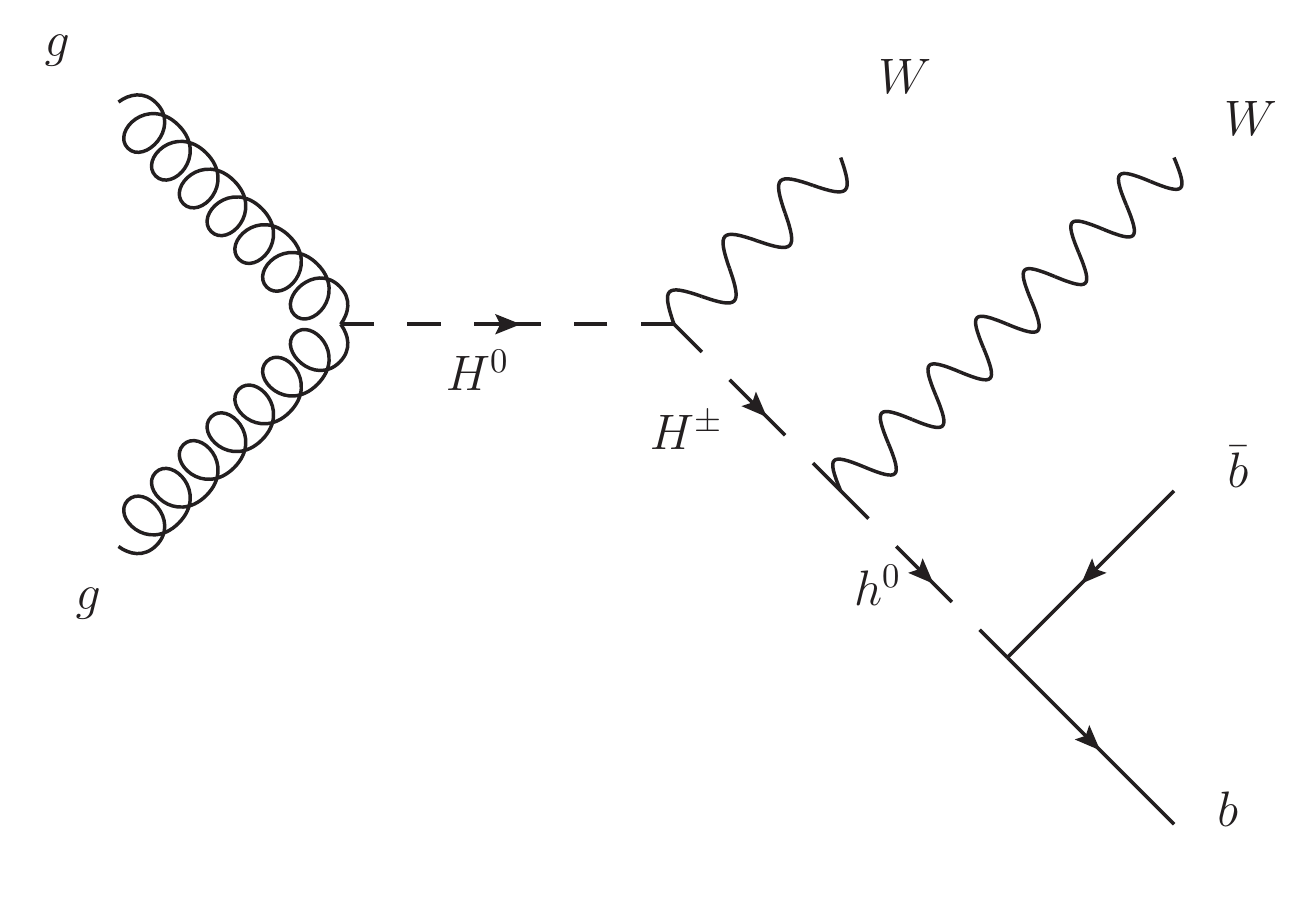}
\end{center}
\caption{ Diagram for $WWb\bar{b}$ production via the cascade
 $gg\rightarrow H^0 \rightarrow H^\pm W^\mp
  \rightarrow WWh^0 \rightarrow WWb\bar{b}$ }
\label{fig:bb}
\end{figure}

In this section, we discuss the search strategy for the cascade
$gg \to H^0 \to H^\pm W^\mp \to W^+ W^- h^0 \to W^+ W^- b\bar{b}$.
%
%
%
Events are reconstructed according to the $t\bar{t}$ hypothesis, in
order to identify and remove this background.  
The neutrino transverse momentum is assumed to be given by the
missing transverse momentum; the longitudinal component is assumed to
the smallest value that gives $(p_\ell+p_\nu)^2=m_W^2$.
The pair of jets without $b$-tags that gives
$m_{jj}$ closest to $m_W$ are labeled as the hadronic $W$ boson decay
products.
In the case that the event contains exactly one $b$-tag,
the leading untagged jet that is not associated to the hadronic $W$ boson
is treated as a second $b$-tagged jet.   
The two $W$ bosons and $b$ jets are paired to form hadronic and
leptonic top quarks according to the assignment that minimizes
$|M_{t}^{\mathrm{lep}} - M_{t}^{\mathrm{had}}|$
where $M_t$ is the $Wb$ invariant mass.

\begin{figure}[h]
\begin{center}
\includegraphics[width=0.48\linewidth]{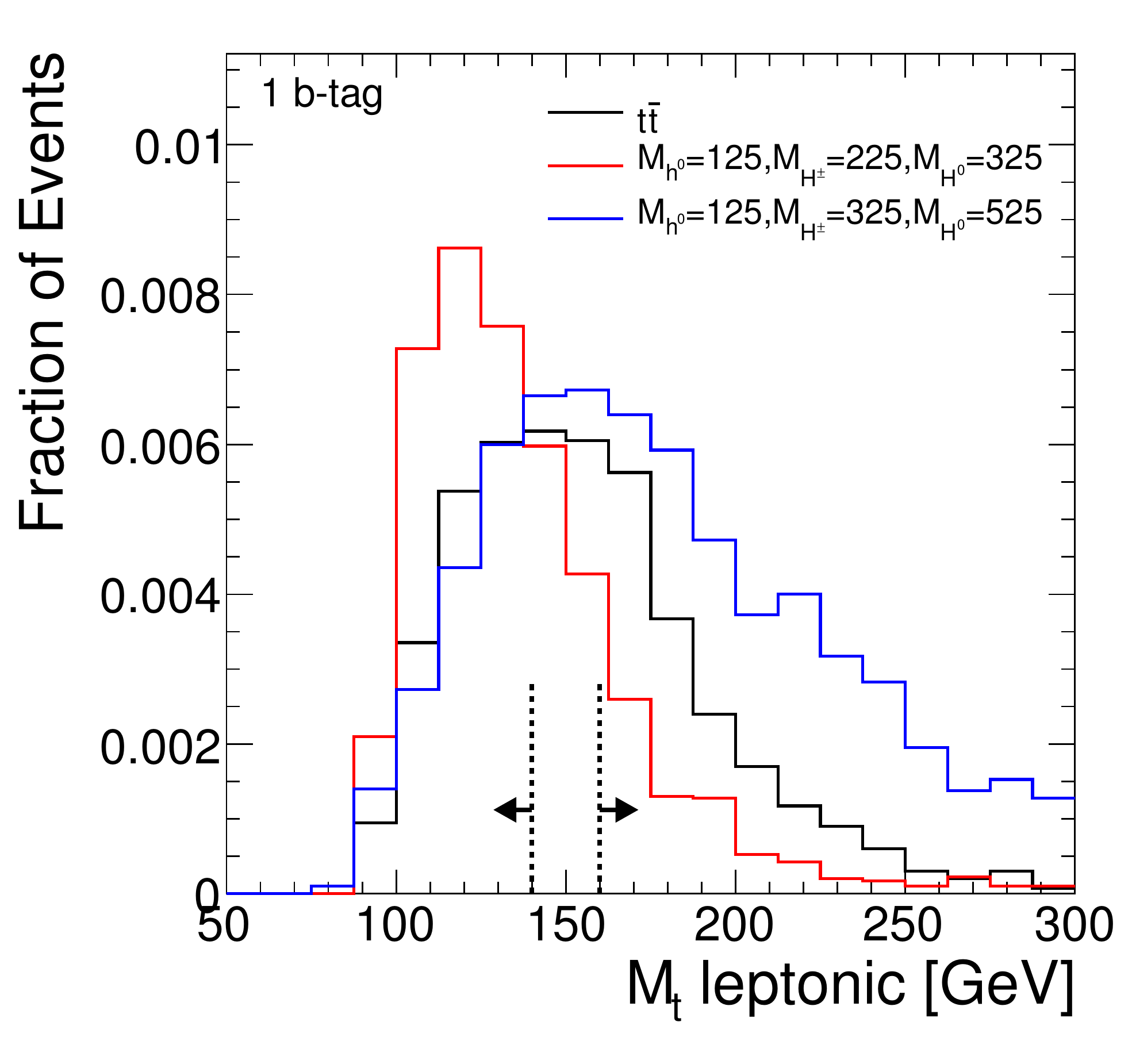}
\includegraphics[width=0.48\linewidth]{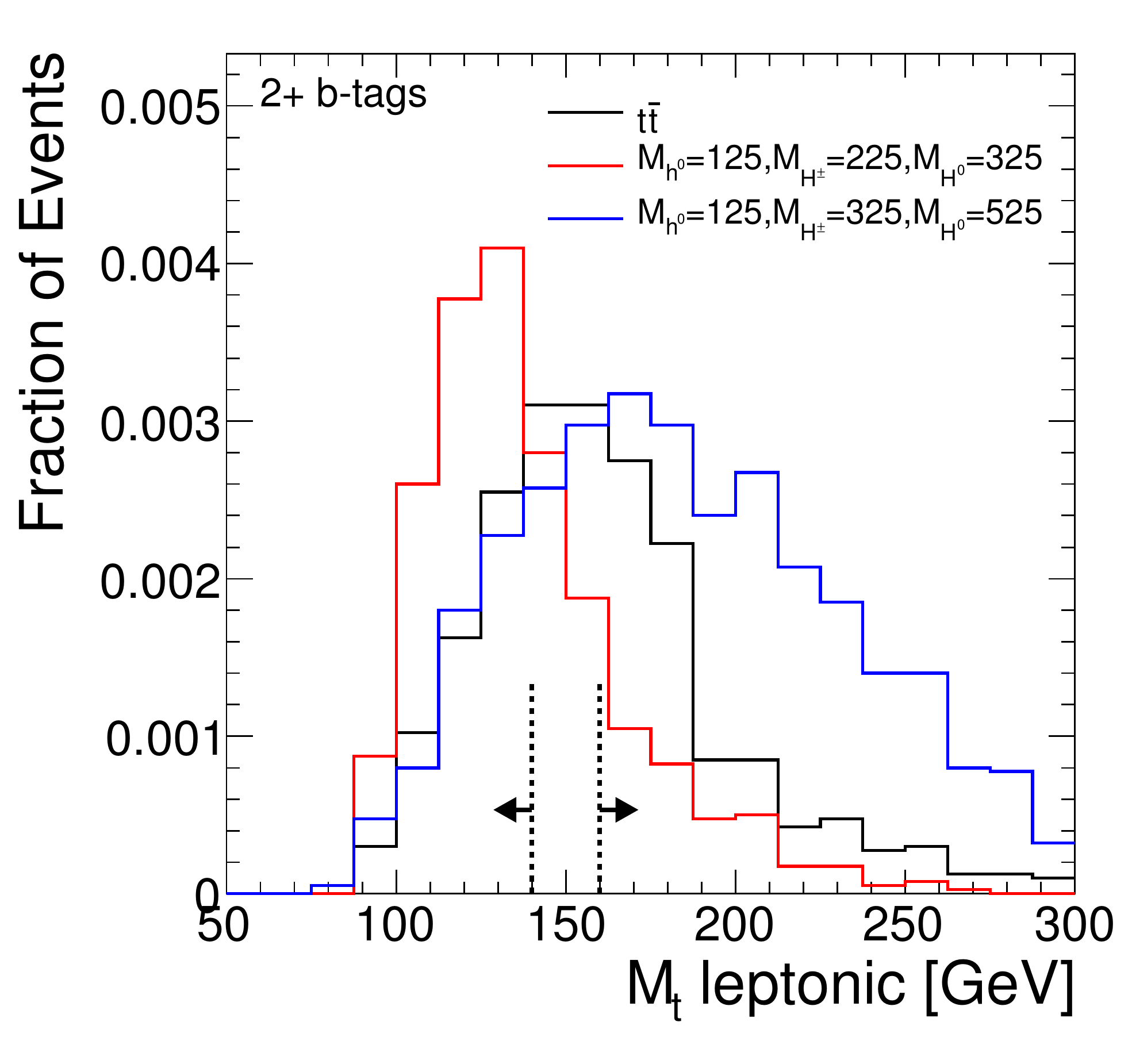}\\
\includegraphics[width=0.48\linewidth]{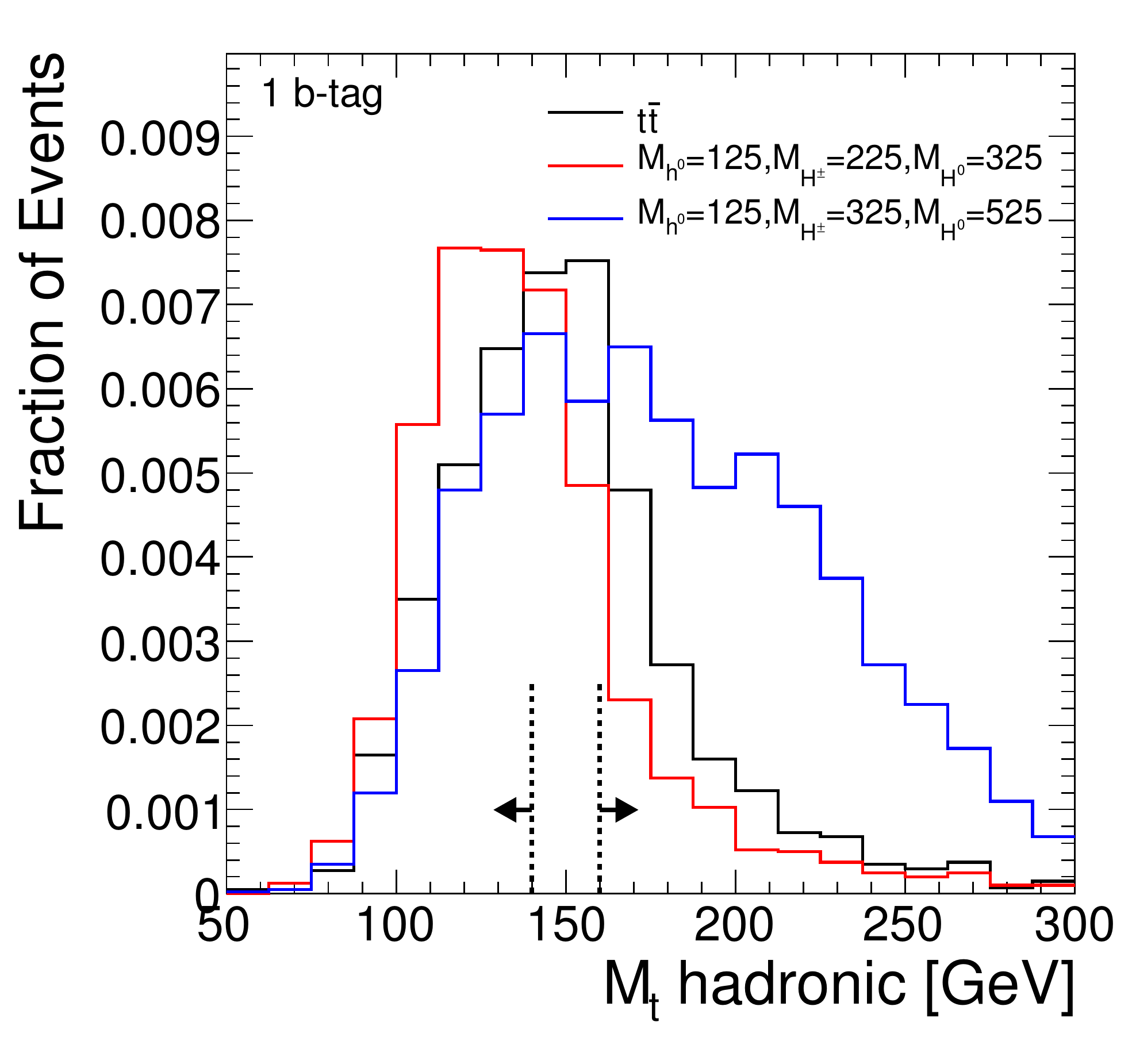}
\includegraphics[width=0.48\linewidth]{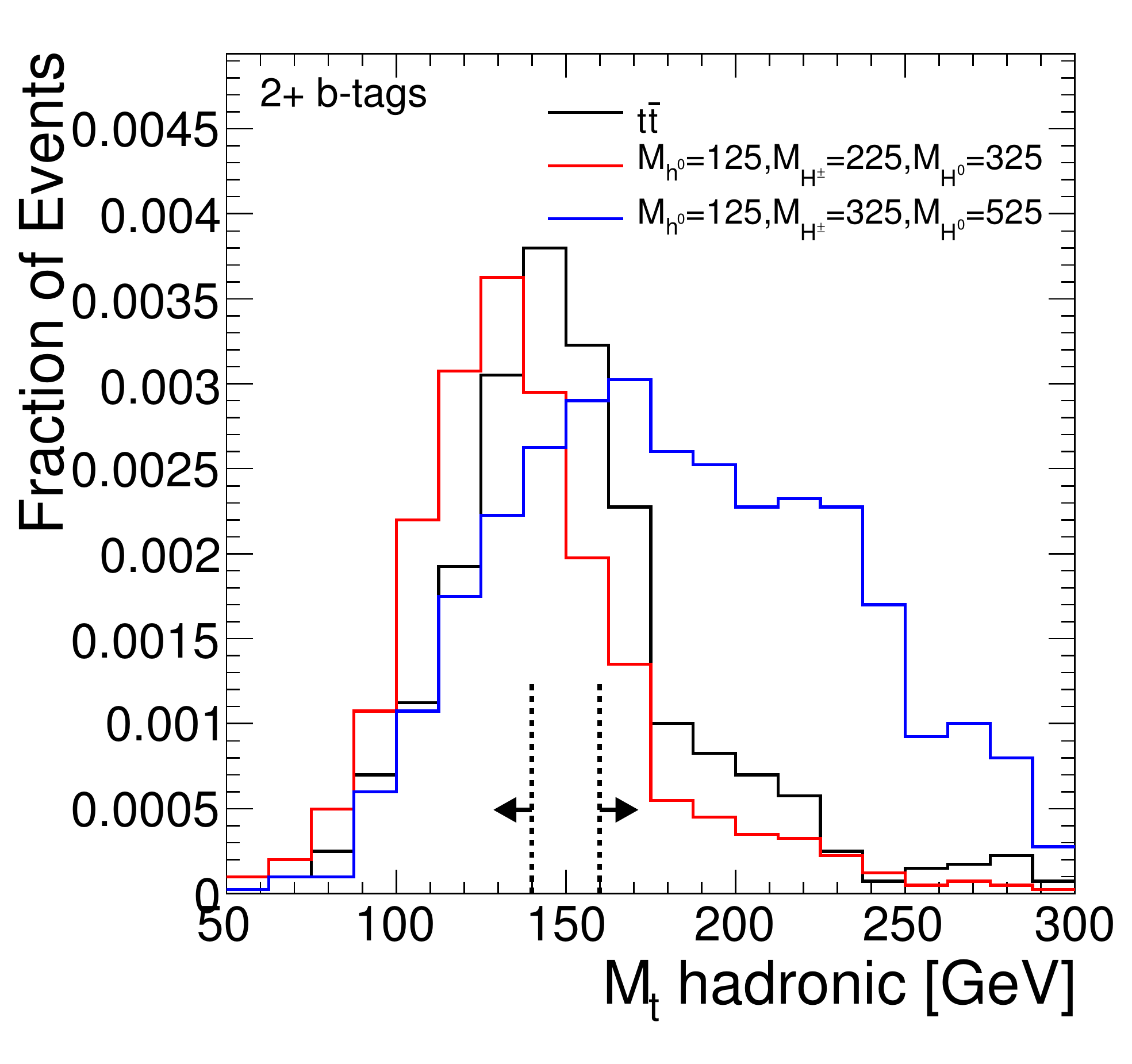}\\
\includegraphics[width=0.48\linewidth]{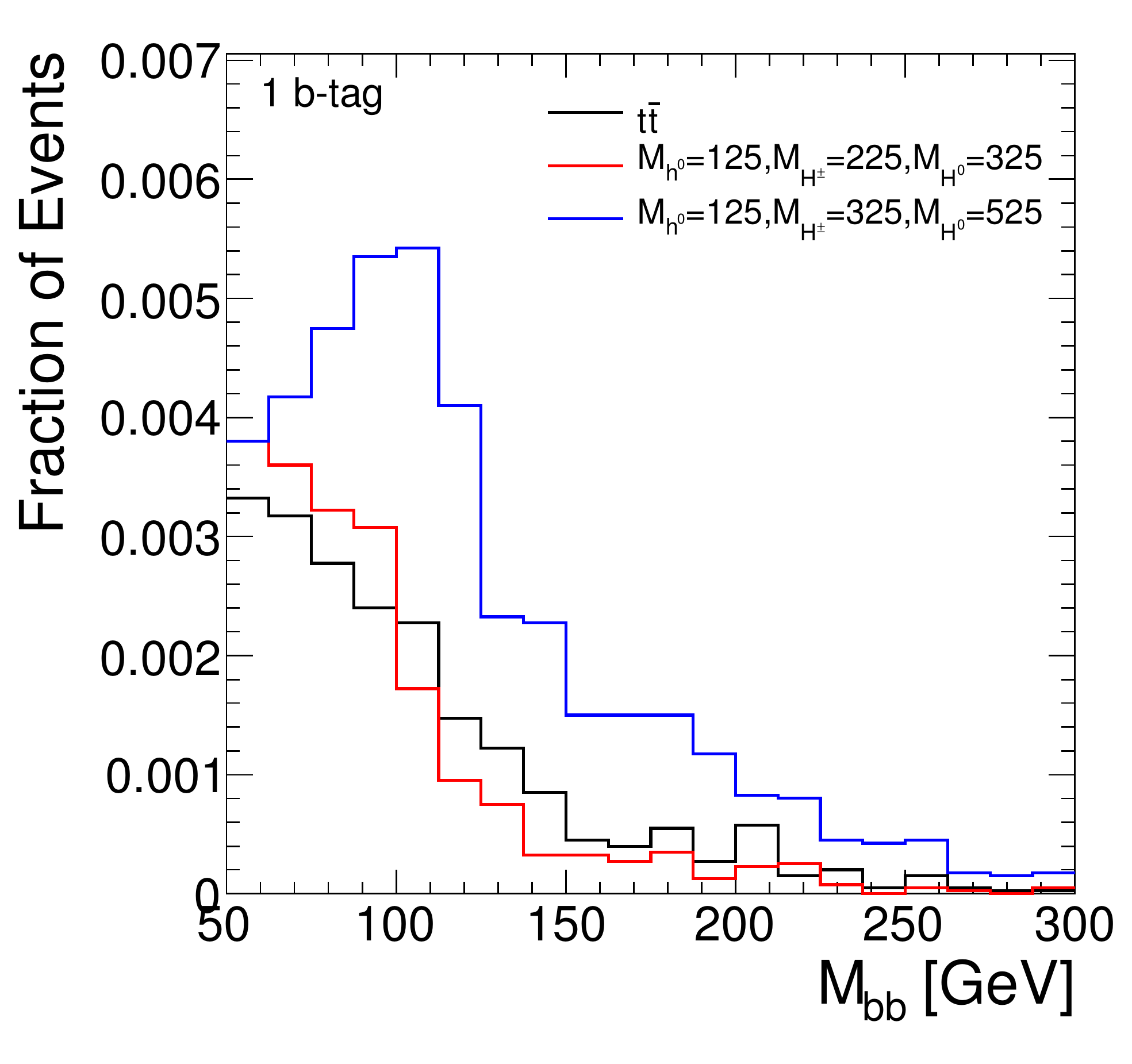}
\includegraphics[width=0.48\linewidth]{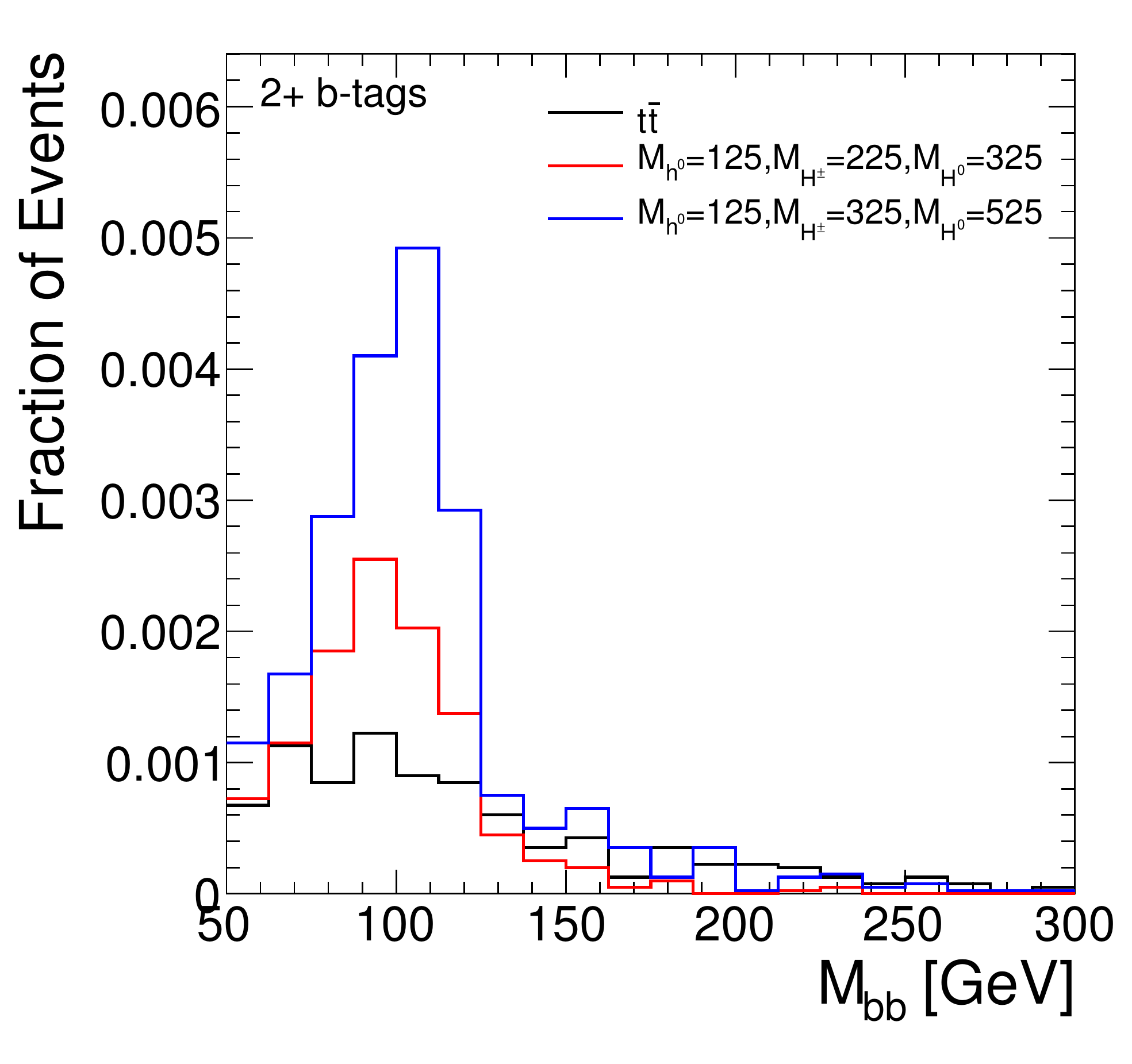}
\end{center}
\caption{ Expected kinematic features from the $W^+W^-h^0 \rightarrow W^+W^-b\bar{b}$ signal shown 
with primary $t\bar{t}$ background at the Tevatron.  Shown are the masses reconstructed as leptonic top (top), 
hadronic top (center) and the $b\bar{b}$ invariant mass (bottom).  Events are categorized
  by the the number of $b$-tags seen: left is exactly one tag, right
  is at least two tags.  The top-quark pair background in the $b\bar{b}$
  mass distribution is suppressed by a top-quark veto, shown in the
  $M_t$ distributions.}
\label{fig:bb_kin}
\end{figure}

The $t\bar{t}$ background shows clear peaks in $M_{t}$
for both leptonic and hadronic modes, see Fig.~\ref{fig:bb_kin}. 
The Higgs cascade decay
lacks the top-quark resonance, leading to reconstructed top-quark masses
further from the top-quark mass.  To reduce the $t\bar{t}$ background, we
veto events if $M_{Wb}^{\mathrm{lep}} \in [ M-10,M+10]$ or
$M_{t}^{\mathrm{had}} \in [ M-10, M+10]$, where $M$ is the median
reconstructed top-quark mass in simulated $t\bar{t}$ events, and the window size
is optimized to maximize expected sensitivity.

For the signal, the $h^0$ mass is
formed by $M_{bb}$ and shows a clear peak in
simulated Higgs cascade events, see Fig.~\ref{fig:bb_kin}.

\begin{figure}[h]
\begin{center}
\includegraphics[width=0.48\linewidth]{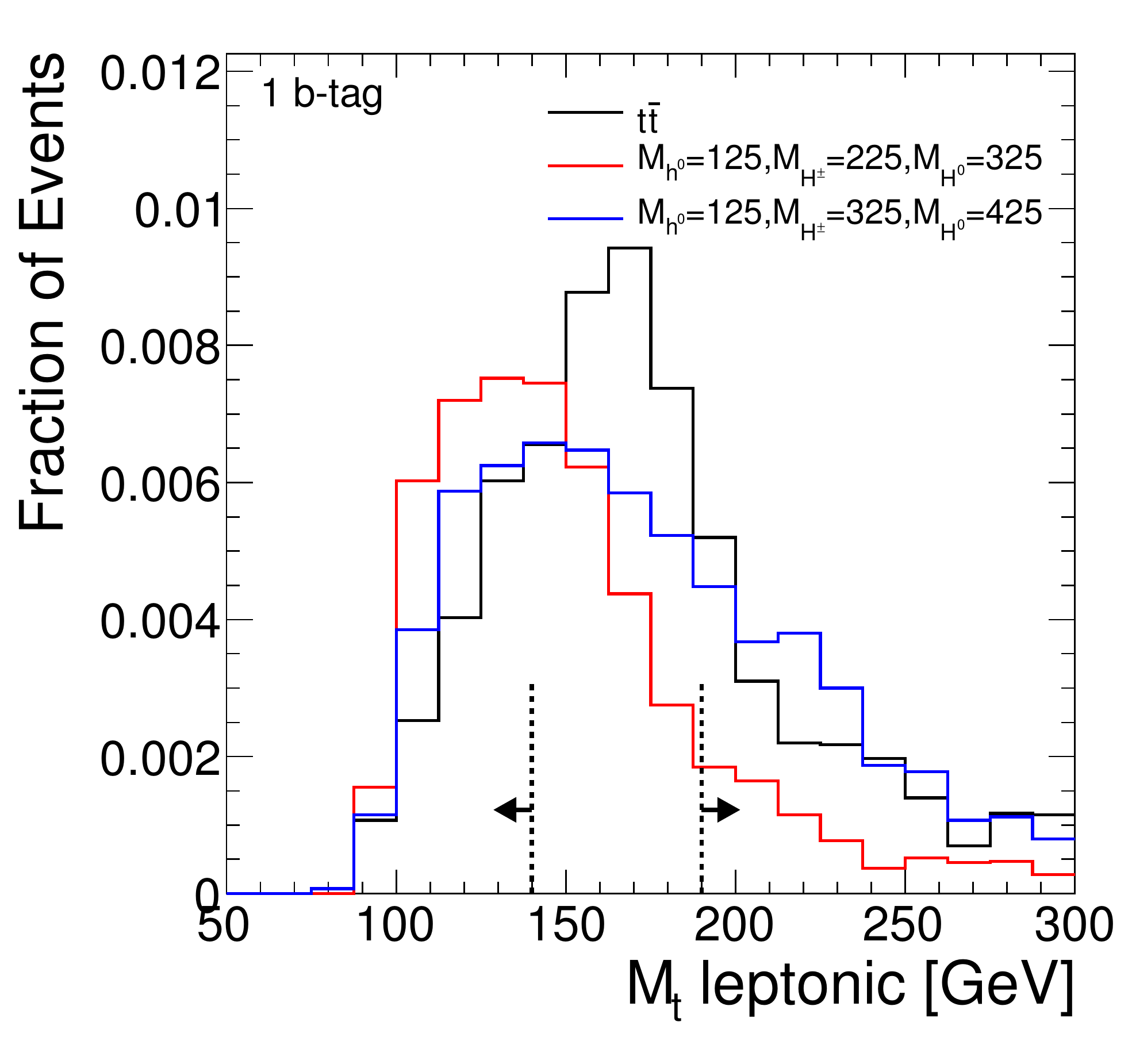}
\includegraphics[width=0.48\linewidth]{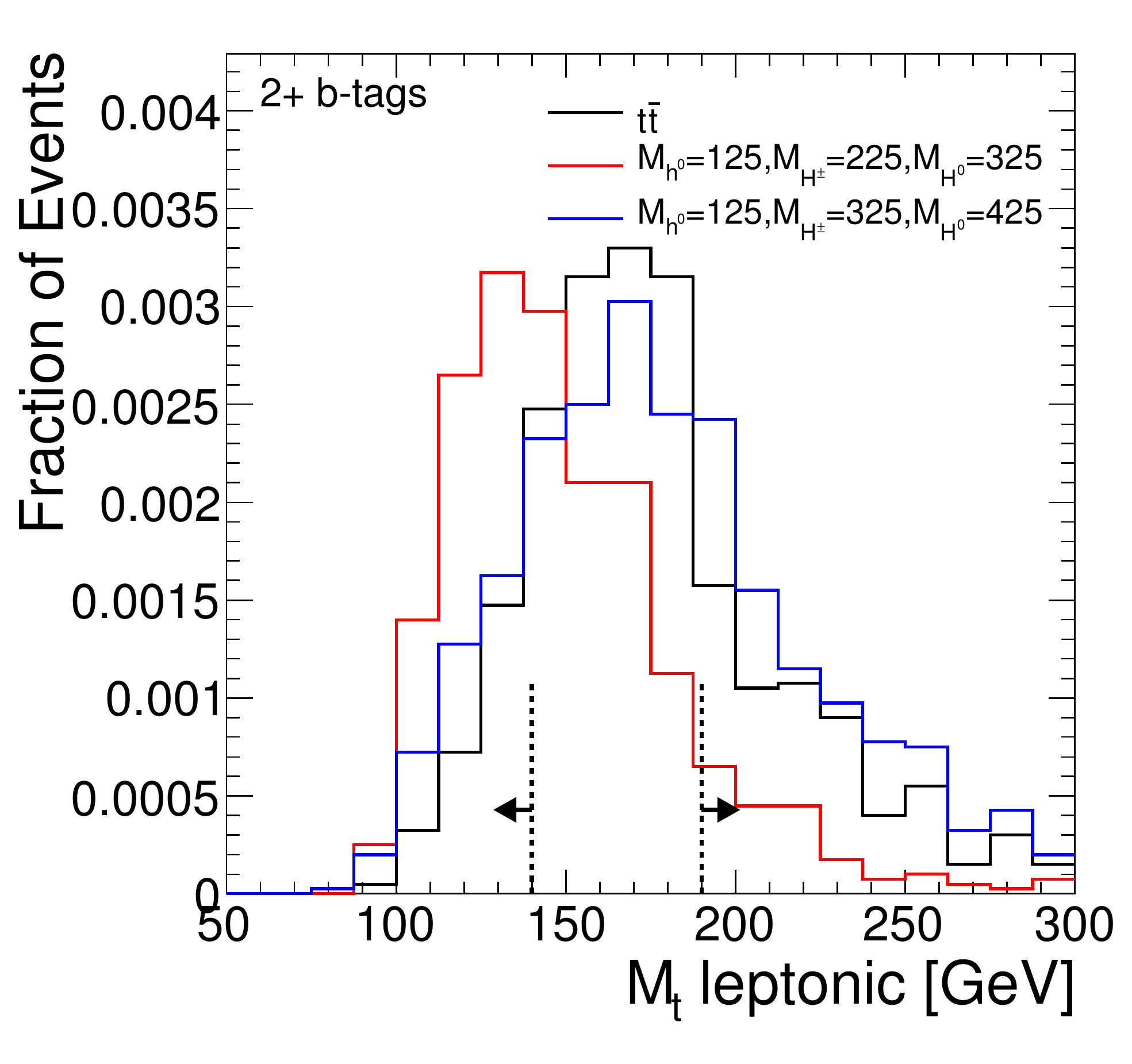}\\
\includegraphics[width=0.48\linewidth]{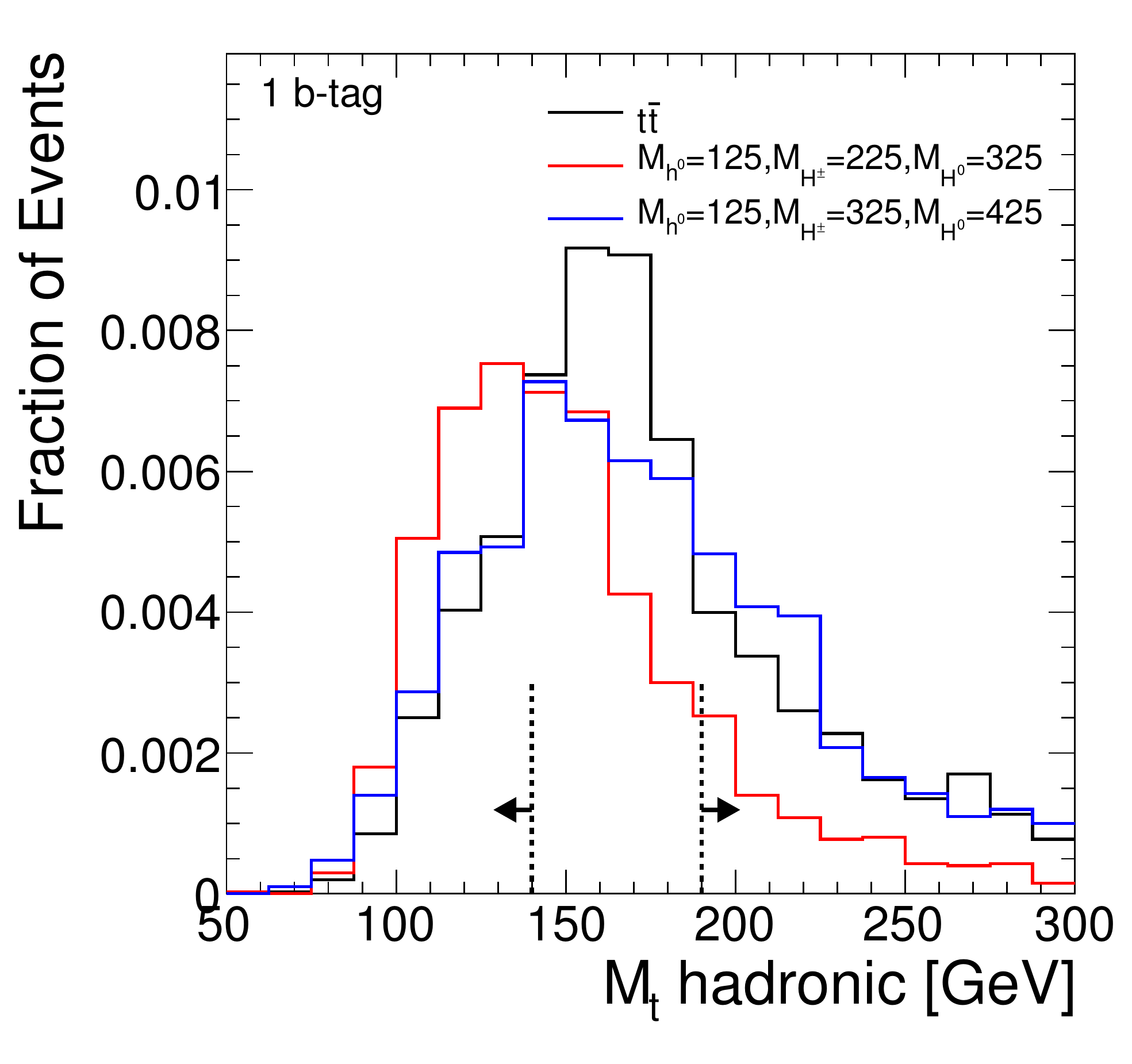}
\includegraphics[width=0.48\linewidth]{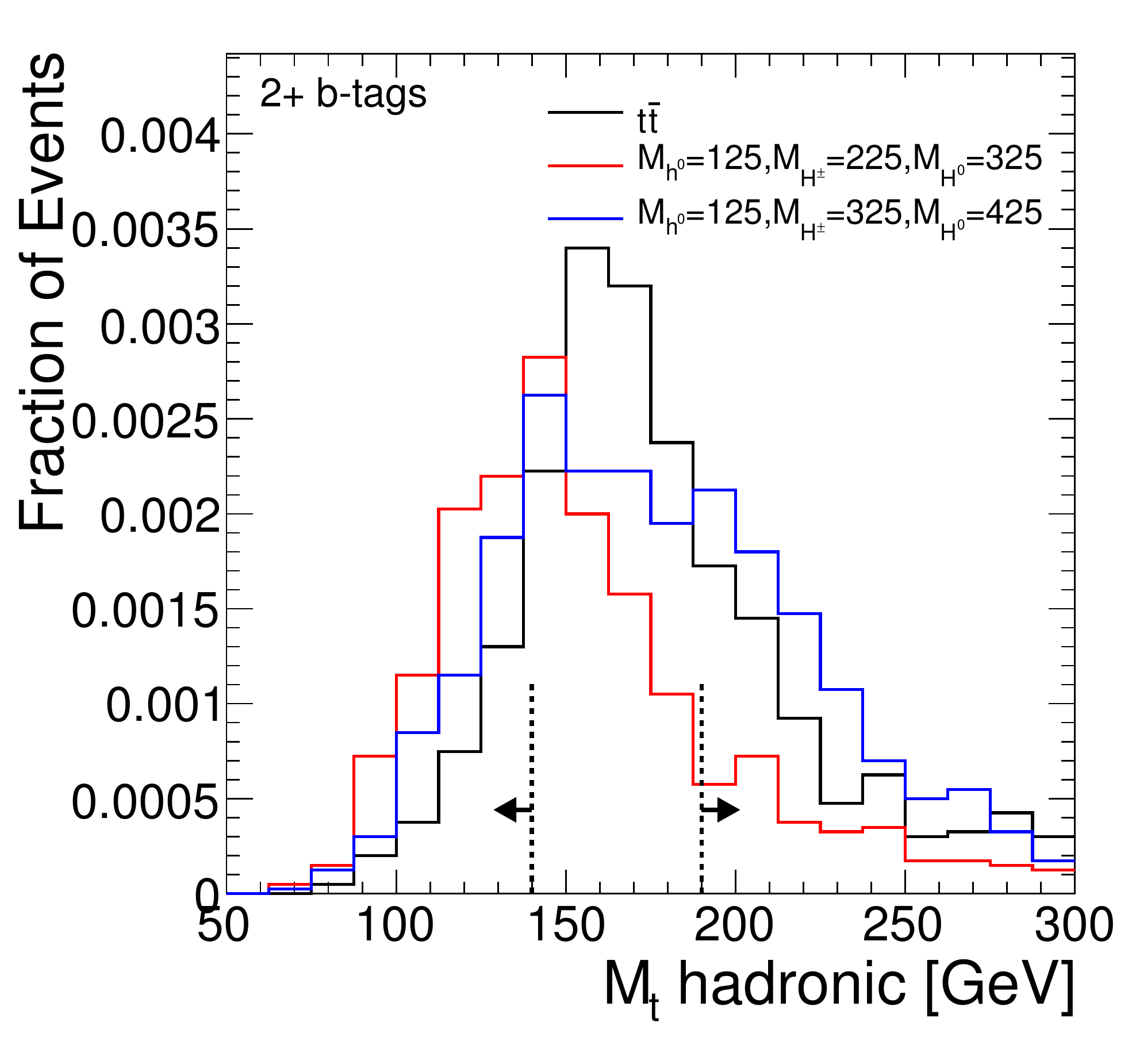}\\
\includegraphics[width=0.48\linewidth]{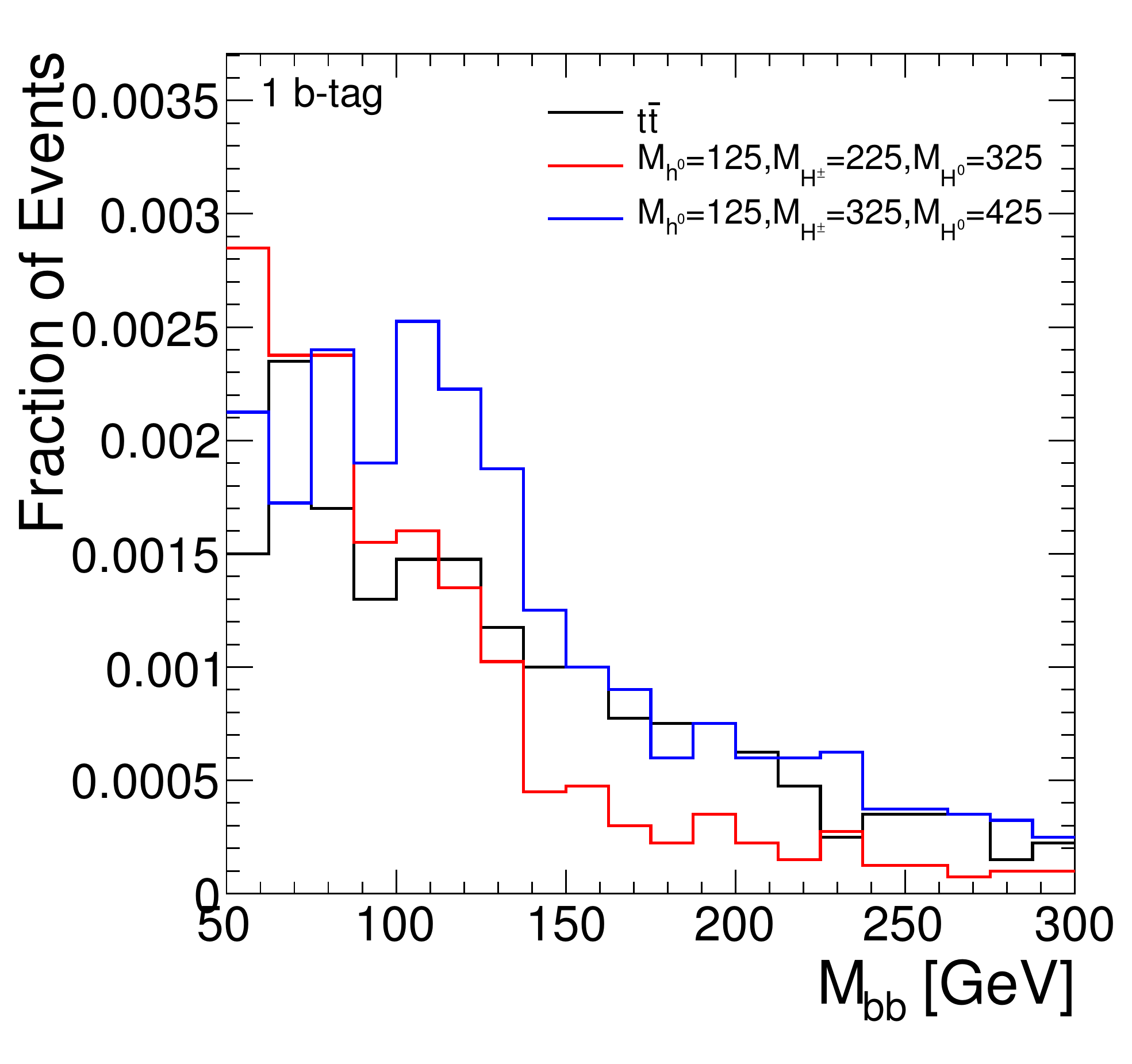}
\includegraphics[width=0.48\linewidth]{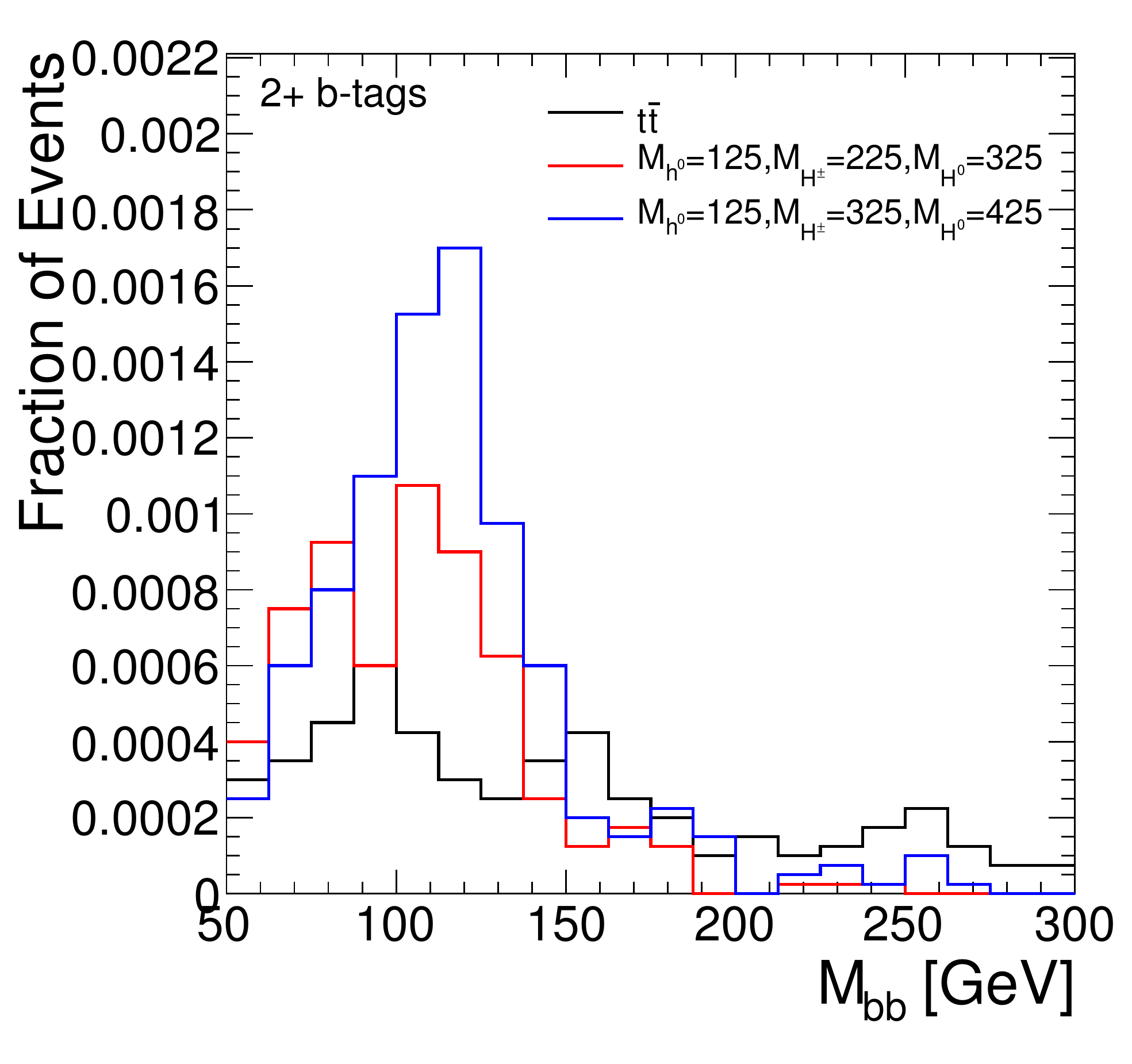}
\end{center}
\caption{ Expected kinematic features from the $W^+W^-h^0 \rightarrow W^+W^-b\bar{b}$ signal shown 
with primary $t\bar{t}$ background at the LHC.  Shown are the masses reconstructed as leptonic top (top), 
hadronic top (center) and the $b\bar{b}$ invariant mass (bottom).  Events are categorized
  by the the number of $b$-tags seen: left is exactly one tag, right
  is at least two tags.  The top-quark pair background in the $b\bar{b}$
  mass distribution is suppressed by a top-quark veto, shown in the
  $M_{t}$ distributions.}
\label{fig:bb_kin_lhc}
\end{figure}

A similar strategy is followed for LHC analysis, see
Fig.~\ref{fig:bb_kin_lhc}, with a wider top-quark veto $[M-25,M+25]$ to suppress the larger top-quark rate at the LHC.

\begin{figure}[h]
\begin{center}
\includegraphics[width=0.48\linewidth]{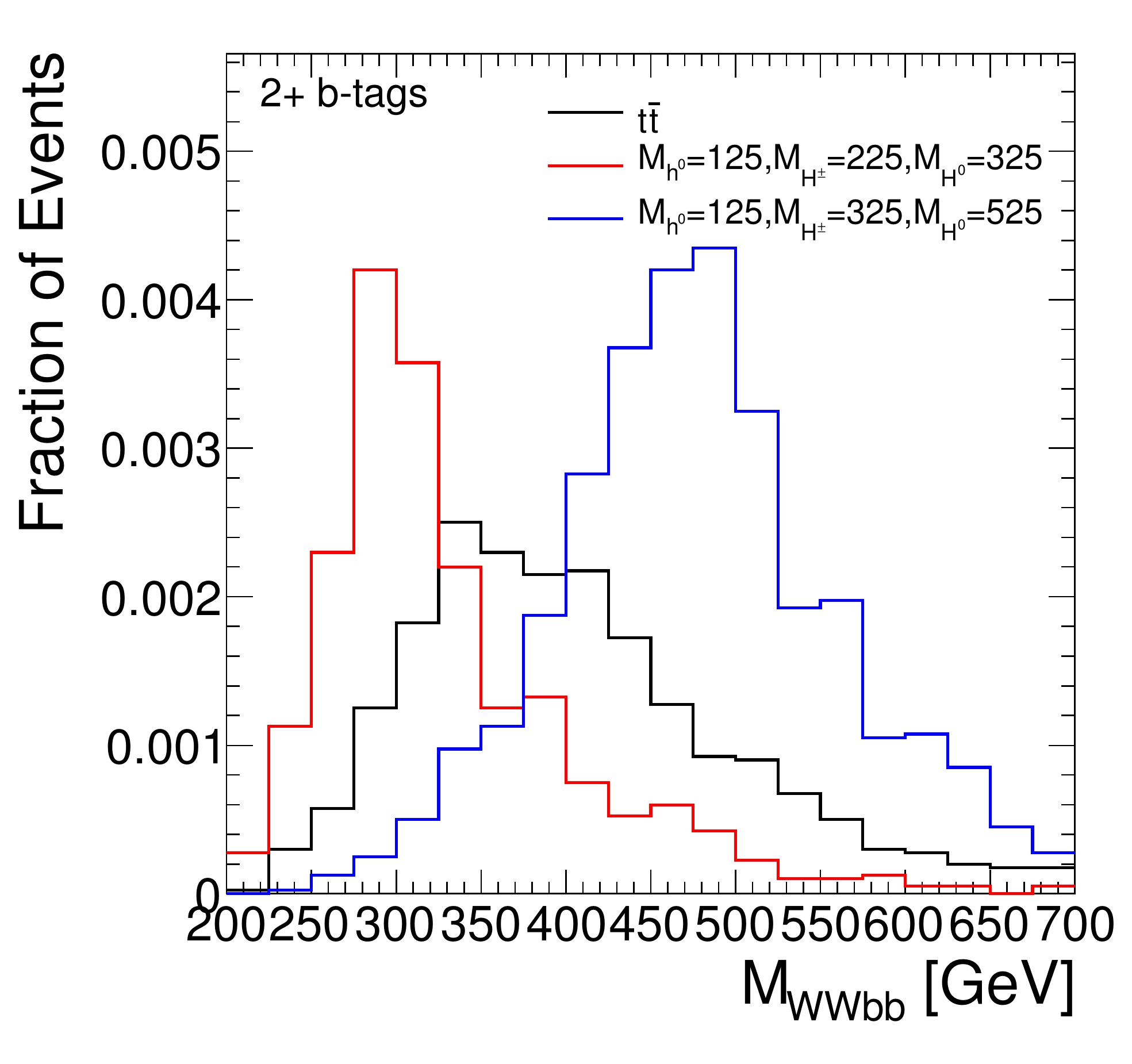}
\includegraphics[width=0.48\linewidth]{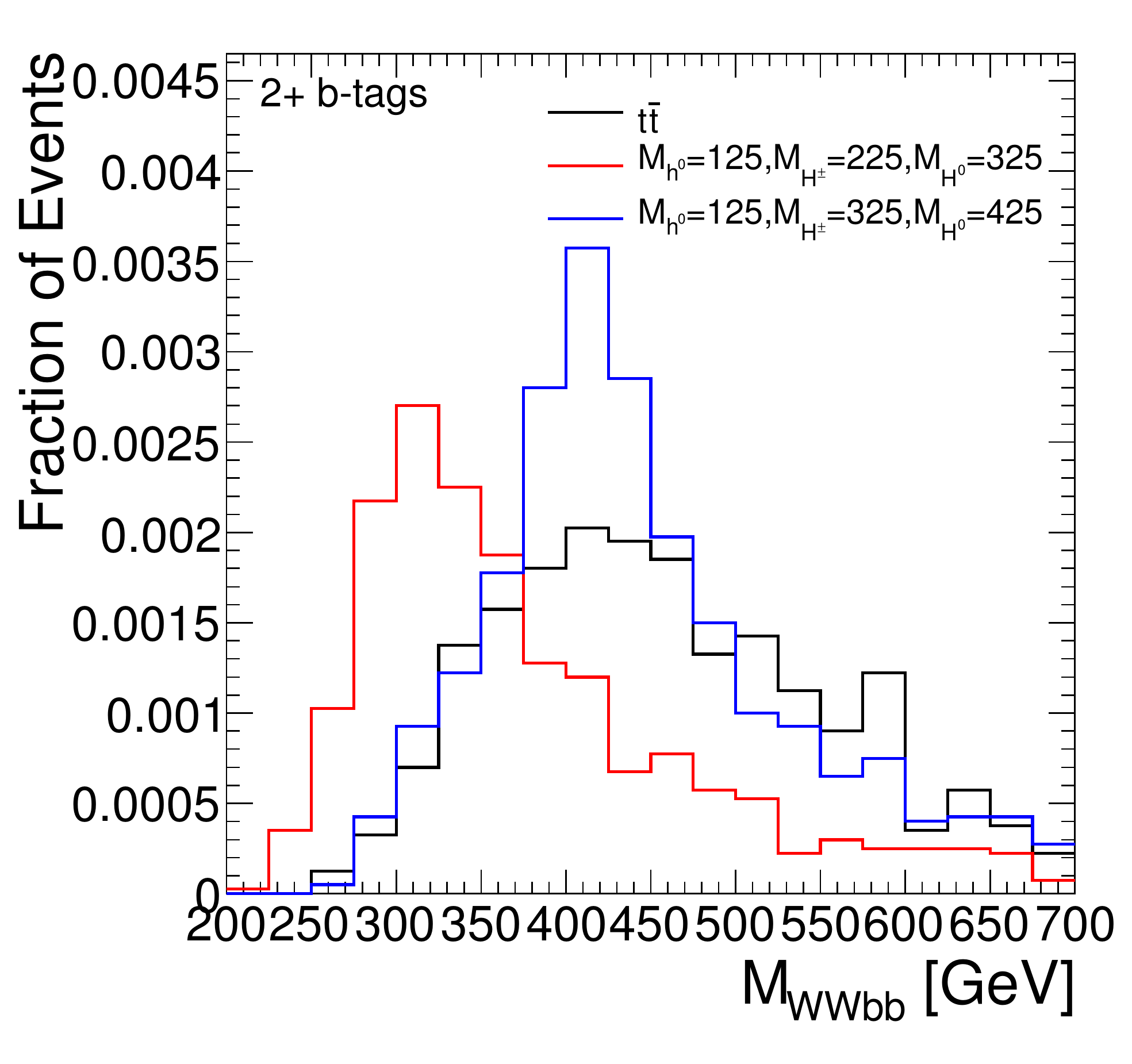}\\
\end{center}
\caption{ Reconstruction of the total invariant mass of the $H^0 \rightarrow H^\pm W^\mp  \rightarrow W^+ W^- h^0 \rightarrow W^+ W^- b\bar{b}$ cascade, as
  $m_{WWbb}$. Left is for the Tevatron; right, LHC.}
\label{fig:bb_mwwbb}
\end{figure}

One could further improve the background rejection by searching for the
$H^{\pm}$ resonance in $m_{Wbb}$, or the $H^0$ resonance in
$m_{WWbb}$, see Fig.~\ref{fig:bb_mwwbb}. The sensitivity gain may be
modest, as the $m_{bb}$ spectrum provides powerful discrimination
between the signal and background shapes while further selection
requirements would reduce the signal acceptance.
A two-dimensional analysis of $m_{bb}$ versus $m_{WWbb}$ is likely
to be most effective.
The simple one-dimensional analysis performed here is sufficient
to demonstrate sensitivity.

\begin{figure}[h]
\begin{center}
\includegraphics[width=0.8\linewidth]{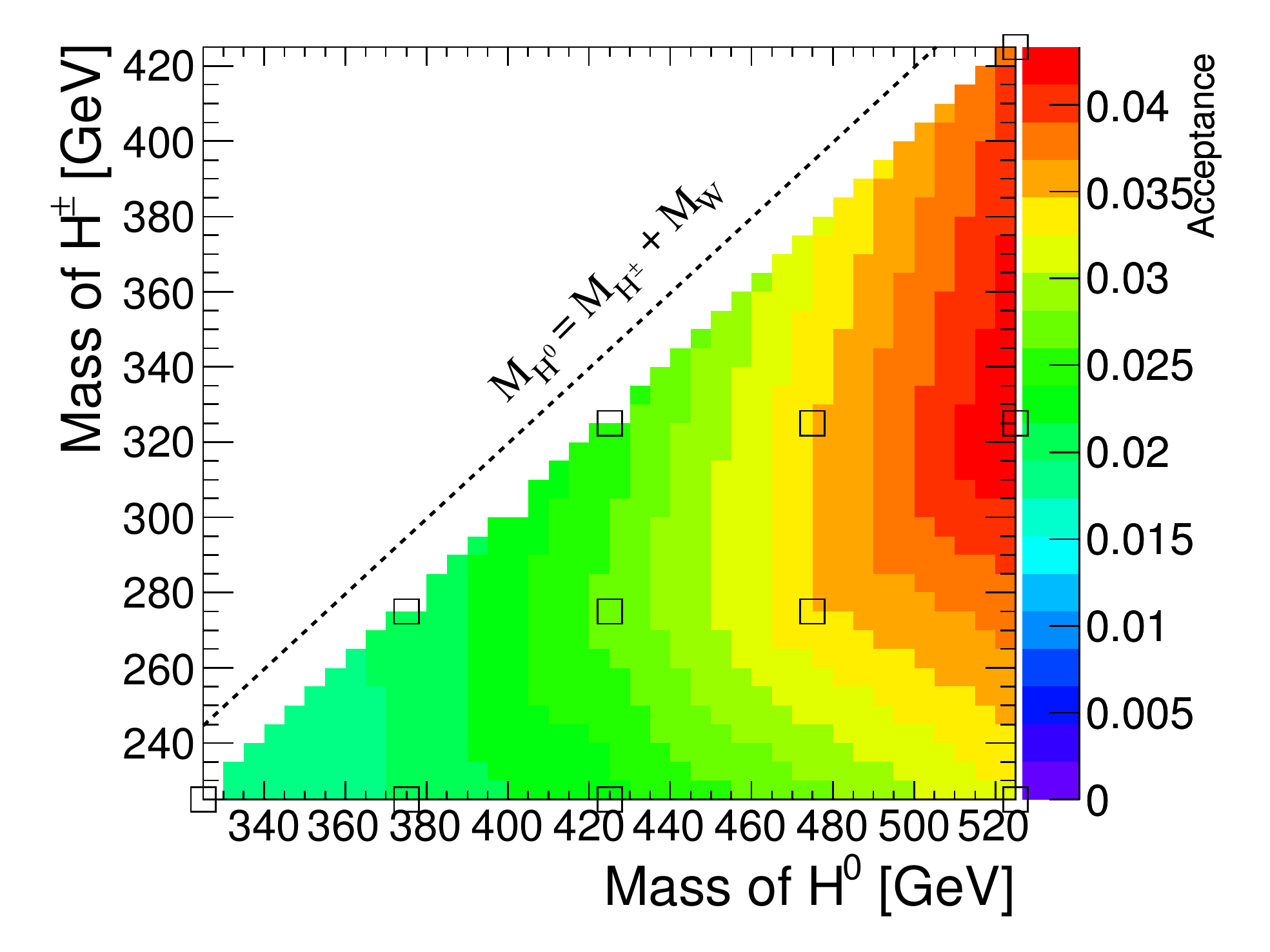}\\
\includegraphics[width=0.8\linewidth]{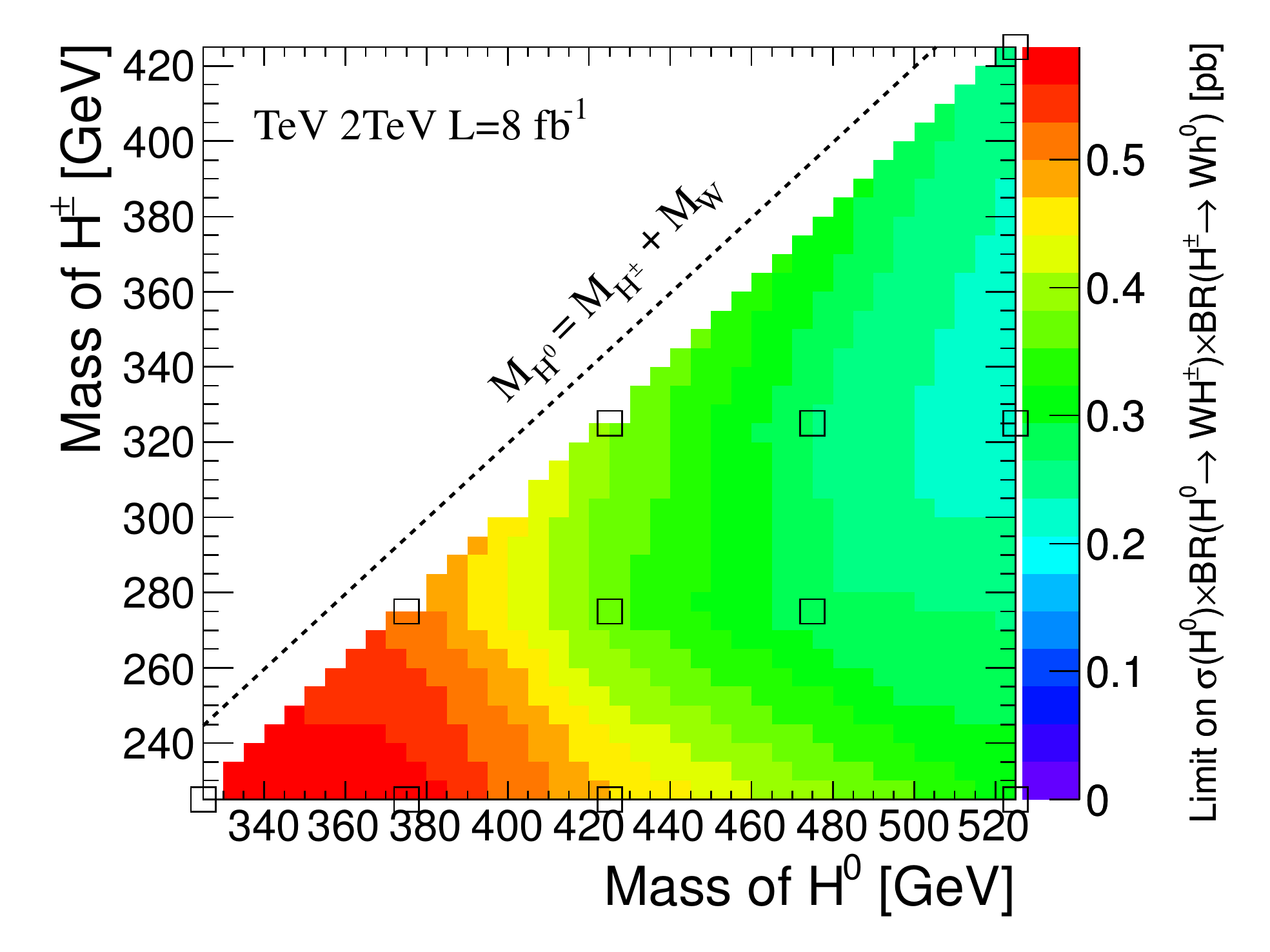}
\end{center}
\caption{ Signal acceptance and expected limits at the Tevatron: top (a) is
$W^+W^-h^0 \rightarrow W^+W^-b\bar{b}$ signal acceptance
after top-quark veto, including the
  branching ratio $WW\rightarrow \ell\nu qq'$; bottom (b) is
  the median
  expected 95\% CL upper limits in the background-only hypothesis.  Both panes use
  $h^0=125$ GeV, and varying $H^0$ and $H^{\pm}$ masses.}
\label{fig:bb_lim}
\end{figure}

\begin{figure}[h]
\begin{center}
\includegraphics[width=0.8\linewidth]{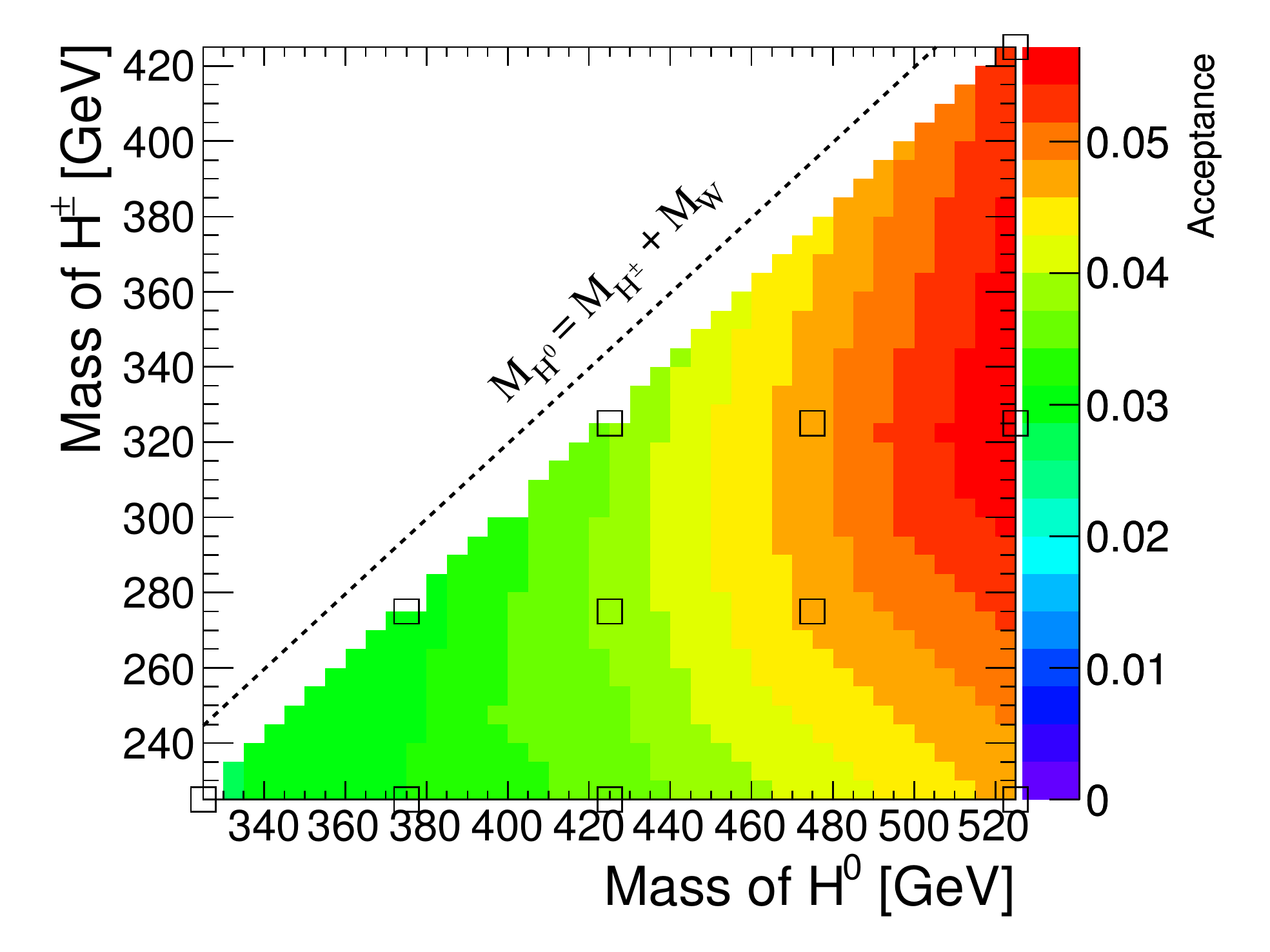}\\
\includegraphics[width=0.8\linewidth]{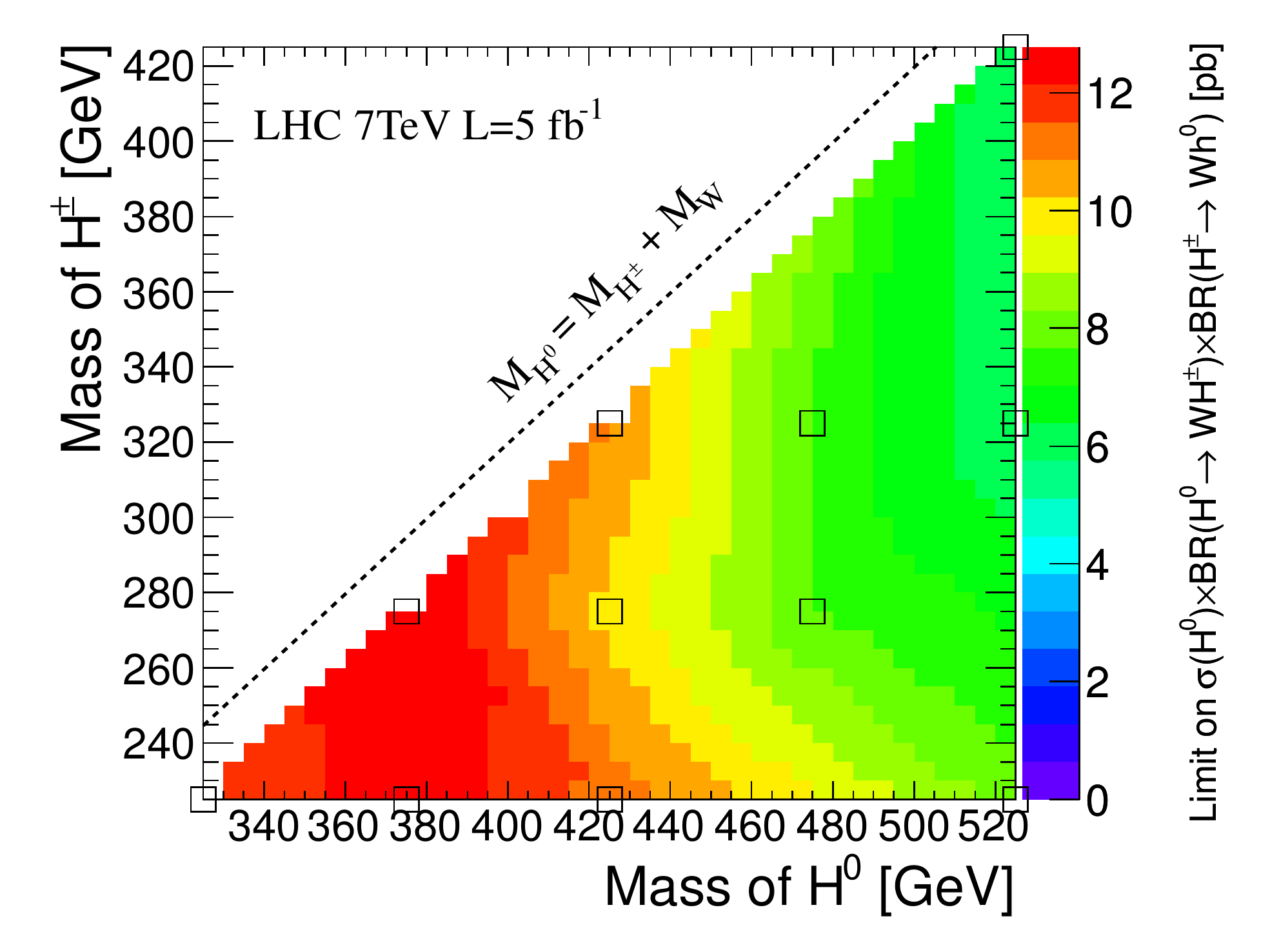}
\end{center}
\caption{ For the LHC: top (a) is $W^+W^-h^0 \rightarrow W^+W^-b\bar{b}$ signal acceptance after top-quark veto, including the
  branching ratio $WW\rightarrow \ell\nu qq'$.  Bottom (b) is median
  expected 95\% CL upper limits in the background-only hypothesis. Both panes use
  $h^0=125$ GeV, and varying $H^0$ and $H^{\pm}$ masses.}
\label{fig:bb_lim_lhc}
\end{figure}

The expected background levels are calculated using the NLO cross section~\cite{ttbarnlo}
for the $t\bar{t}$, acceptance calculated with simulated events, and a
luminosity of 8~fb$^{-1}$ (5~fb$^{-1}$) for the Tevatron (LHC).
A 10\% uncertainty is assumed.  
The signal acceptance is calculated using
simulated events, see Figs.~\ref{fig:bb_lim}a \& \ref{fig:bb_lim_lhc}a.

To extract the most likely value of the signal cross section, we
analyze the shape of the $m_{bb}$ distribution. Specifically, a binned
maximum likelihood fit is used in the $m_{bb}$ variable, floating each background rate within
uncertainties, allowing variation due to systematic uncertainties described above.  The signal and background rates are
fit simultaneously.  The top-quark mass veto described above improves
the sensitivity by approximately 20\%, by significantly reducing the
background efficiency relative to the signal efficiency. The CLs method~\cite{cls} is used to set 95\%
cross section upper limits.  The median expected upper limit is
extracted in the background-only hypothesis, see Figs.~\ref{fig:bb_lim}b \& \ref{fig:bb_lim_lhc}b.

\section{Resonances in $Wbb$}

\begin{figure}[h]
\begin{center}
\includegraphics[width=0.8\linewidth]{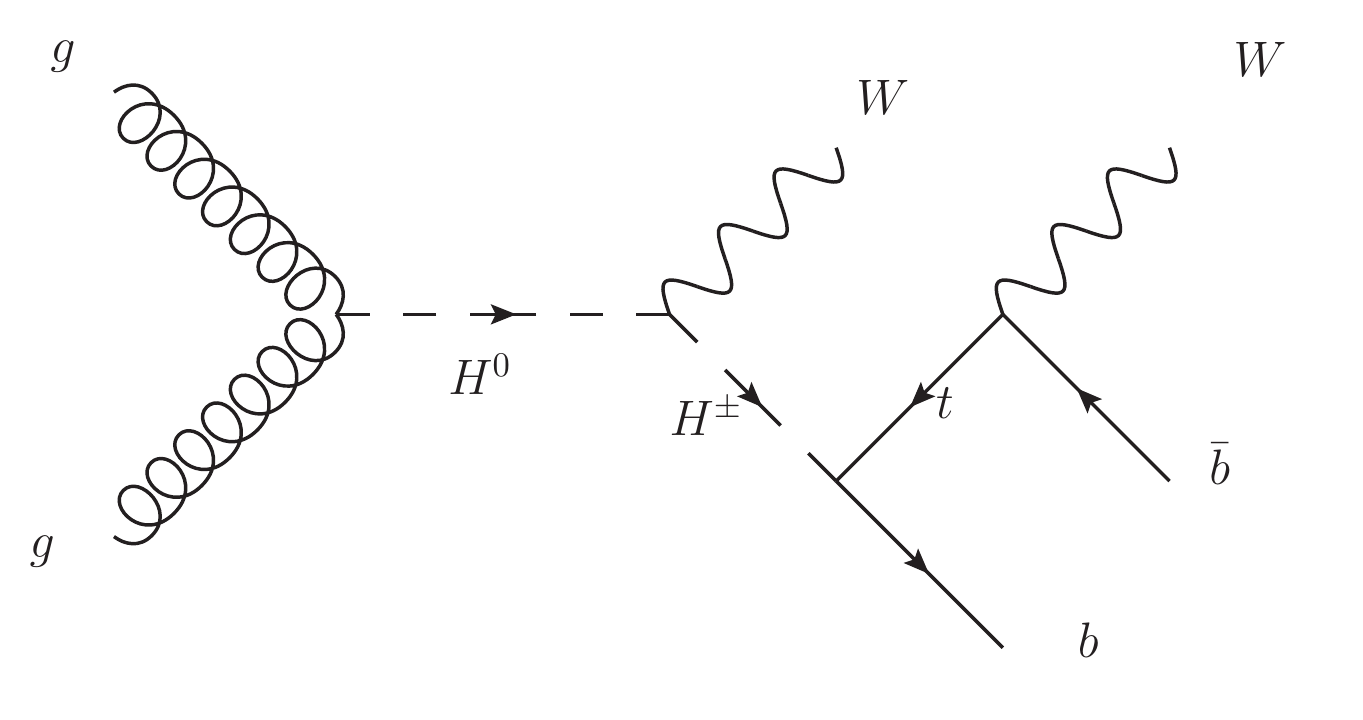}
\end{center}
\caption{ Diagram for $W^+W^-b\bar{b}$ production via the cascade $gg\rightarrow H^0 \rightarrow H^\pm W^\mp
  \rightarrow Wtb \rightarrow W^+W^-b\bar{b}$. }
\label{fig:wbb}
\end{figure}

Production of $W^+W^-b\bar{b}$ may also occur through the cascade process (Fig.~\ref{fig:wbb}),

\[ gg\rightarrow H^0 \rightarrow H^\pm W^\mp
  \rightarrow Wtb \rightarrow W^+W^-b\bar{b} \]

\noindent
which gives resonant production of $H^{\pm}\rightarrow W^\pm b\bar{b}$, and
kinematics distinct from $t\bar{t}\rightarrow W^+W^-b\bar{b}$.

\begin{figure}[t]
\begin{center}
\includegraphics[width=0.48\linewidth]{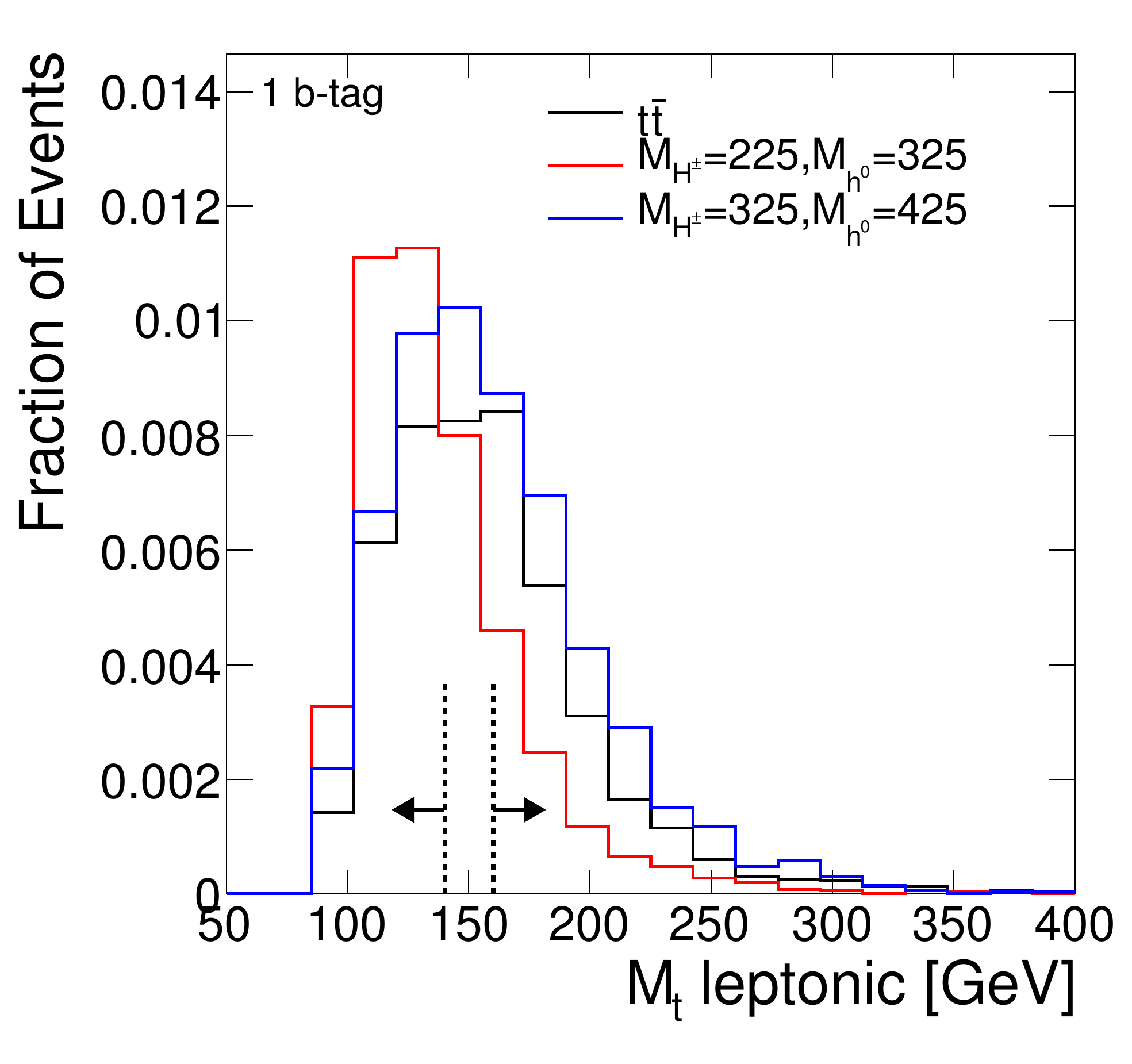}
\includegraphics[width=0.48\linewidth]{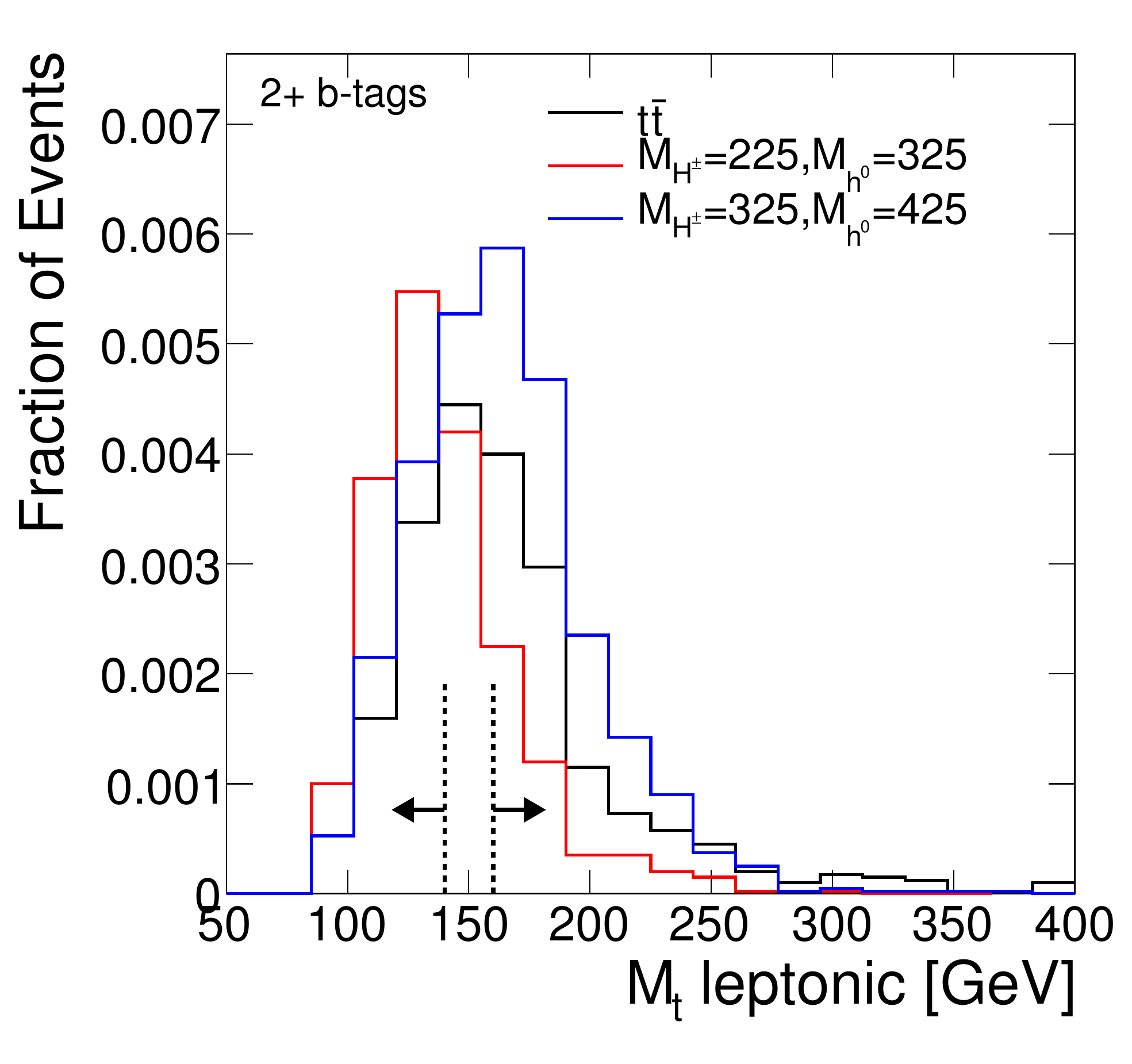}\\
\includegraphics[width=0.48\linewidth]{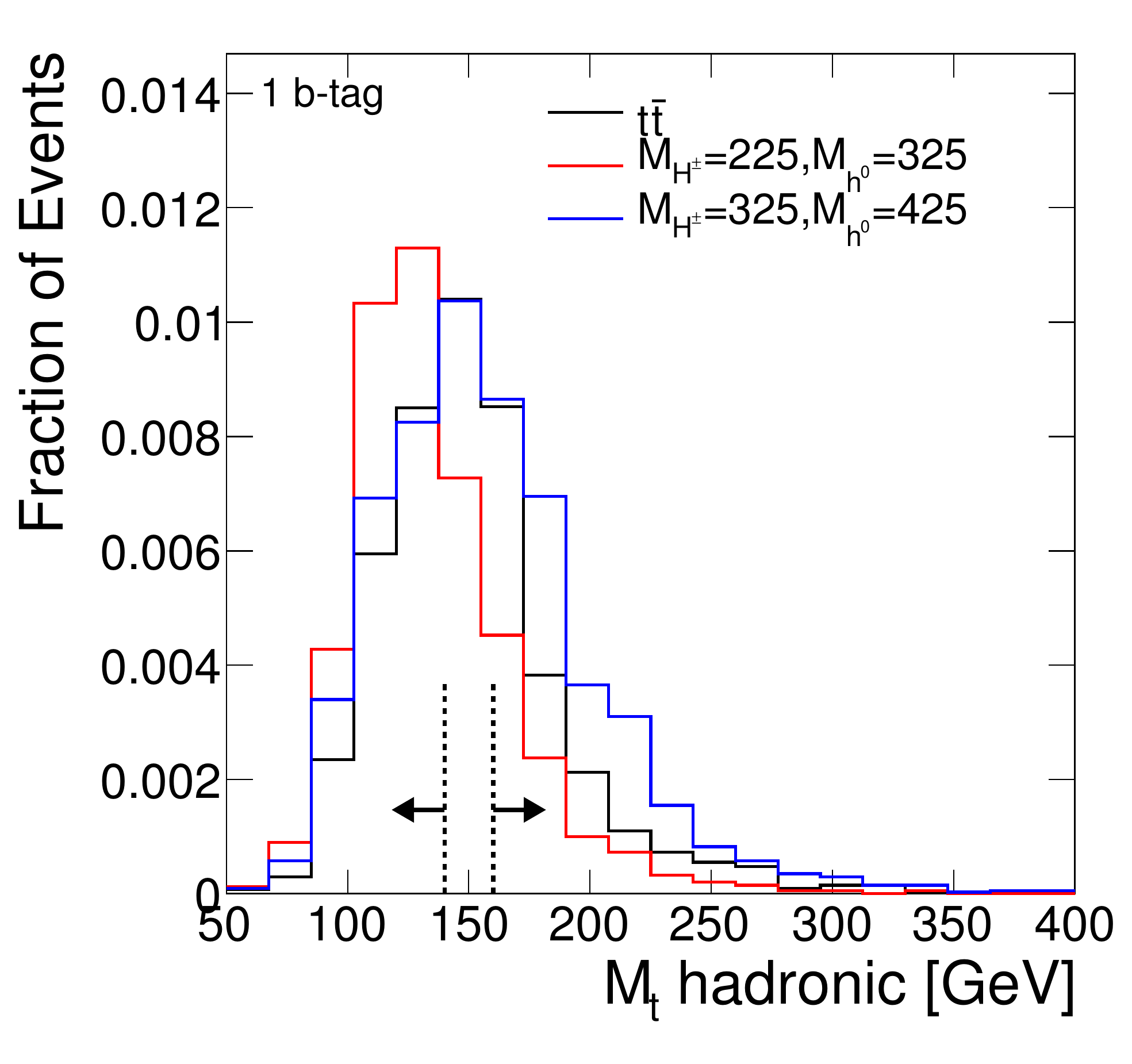}
\includegraphics[width=0.48\linewidth]{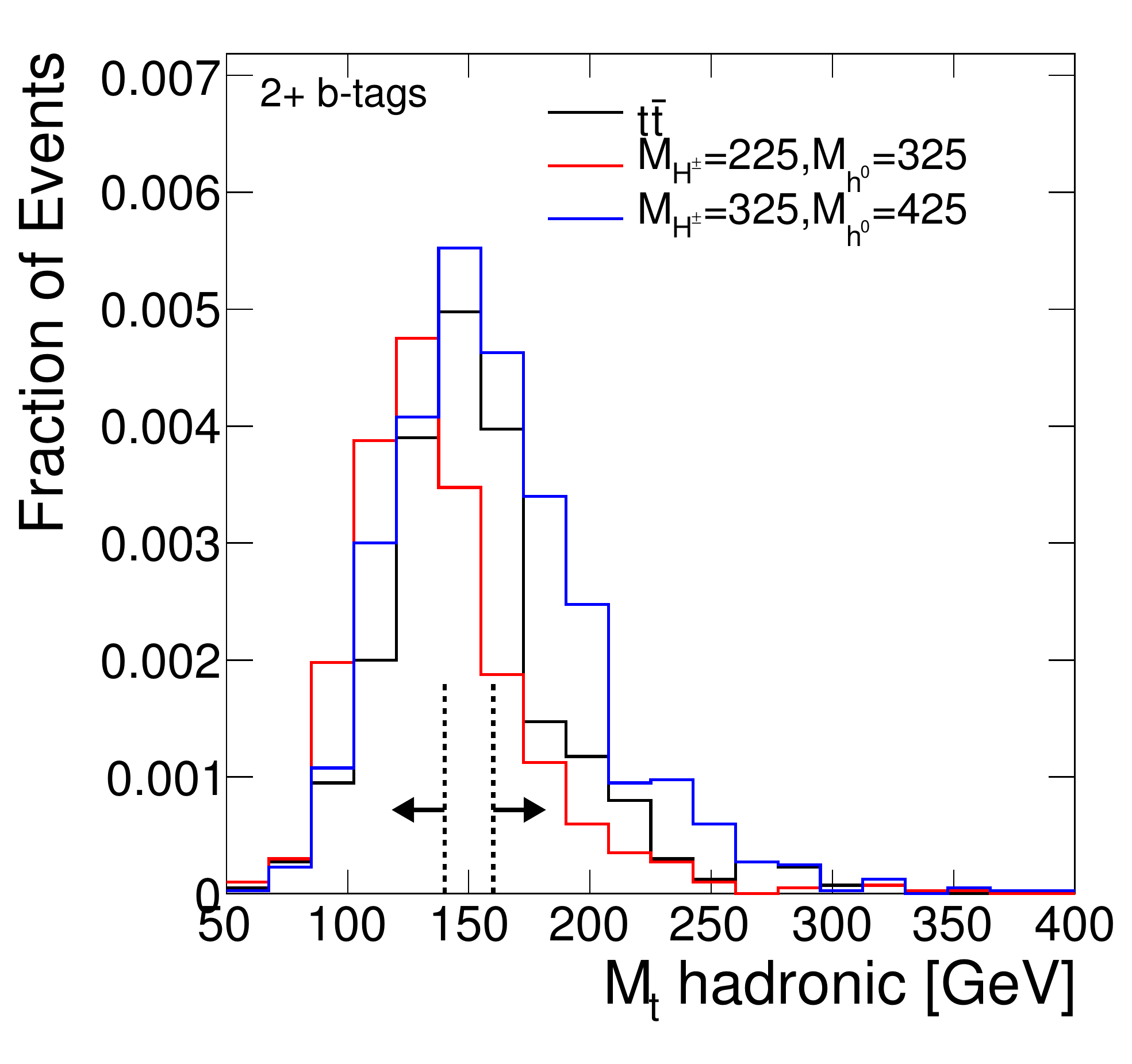}\\
\includegraphics[width=0.48\linewidth]{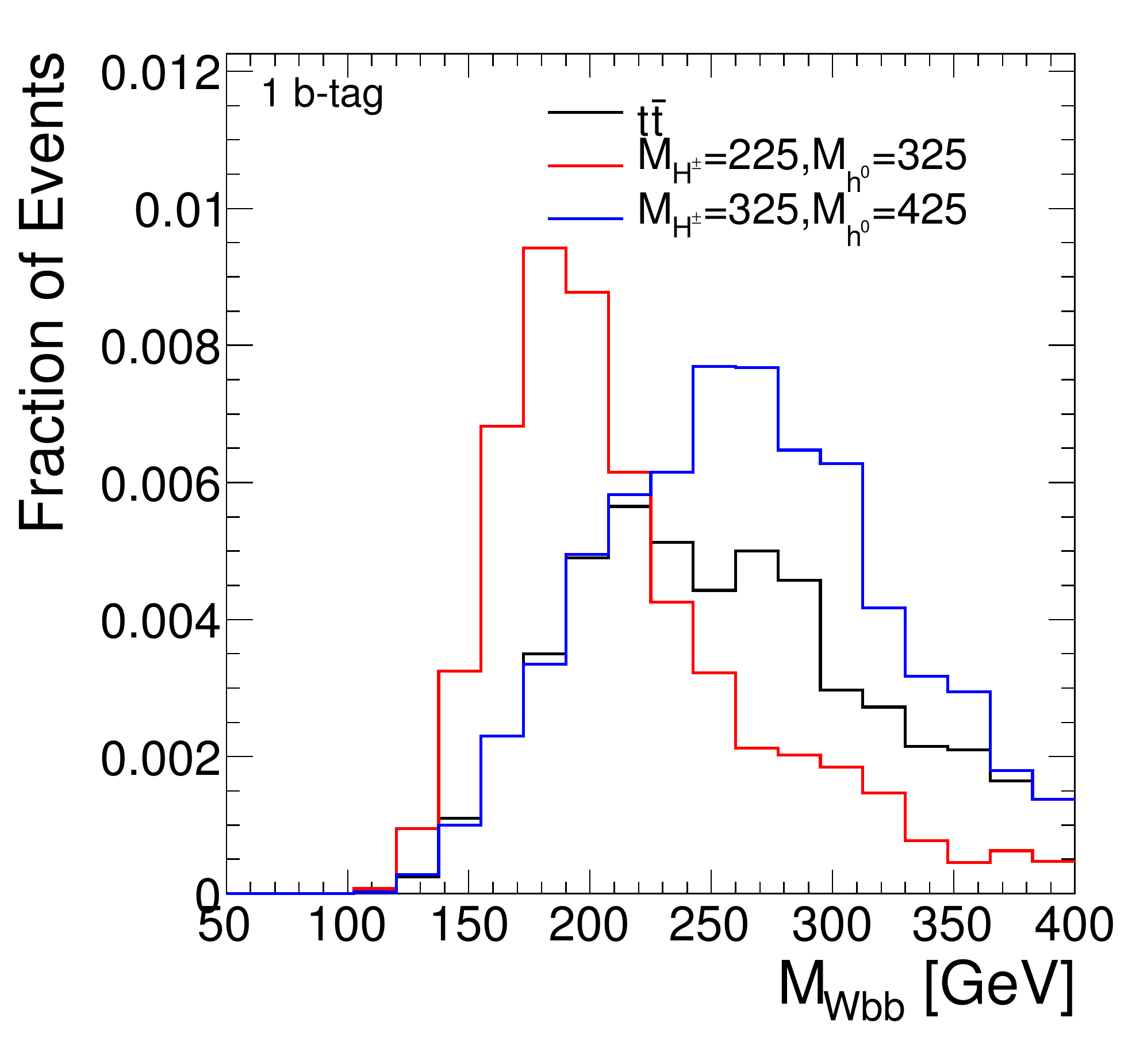}
\includegraphics[width=0.48\linewidth]{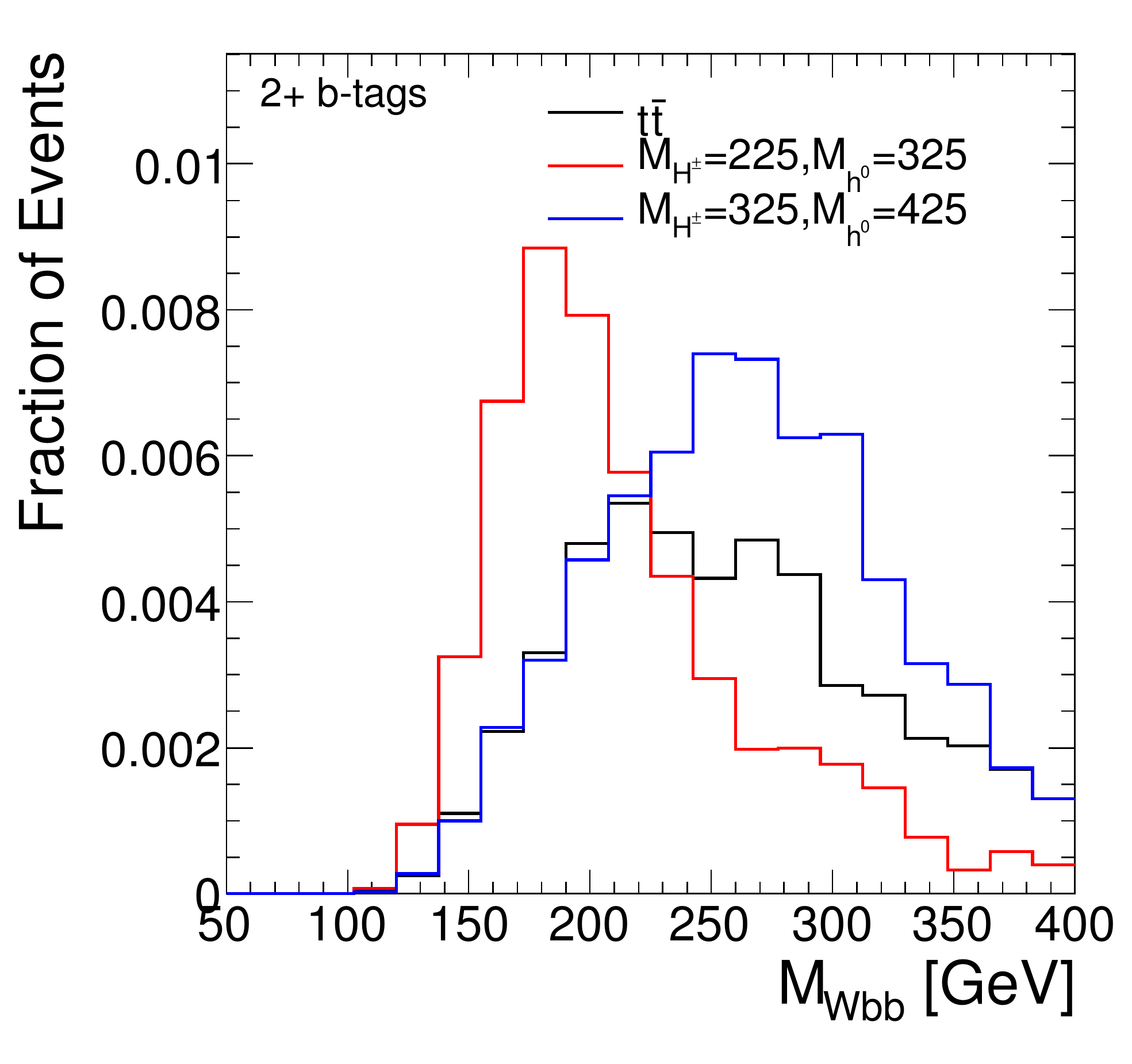}
\end{center}
\caption{ Expected kinematic features of $H^{\pm}\rightarrow W^\pm b\bar{b}$ resonance signal and background events at the
  Tevatron.  Shown are leptonic top mass (top), hadronic top mass
  (center) and $Wb\bar{b}$ invariant mass (bottom).  Events are categorized
  by the the number of $b$-tags seen: left is exactly one tag, right
  is at least two tags.  The top-quark pair background in the $Wb\bar{b}$
  mass distribution is suppressed by a top-quark veto, shown in the
  $M_t$ distributions.}
\label{fig:wbb_kin}
\end{figure}

\begin{figure}[t]
\begin{center}
\includegraphics[width=0.48\linewidth]{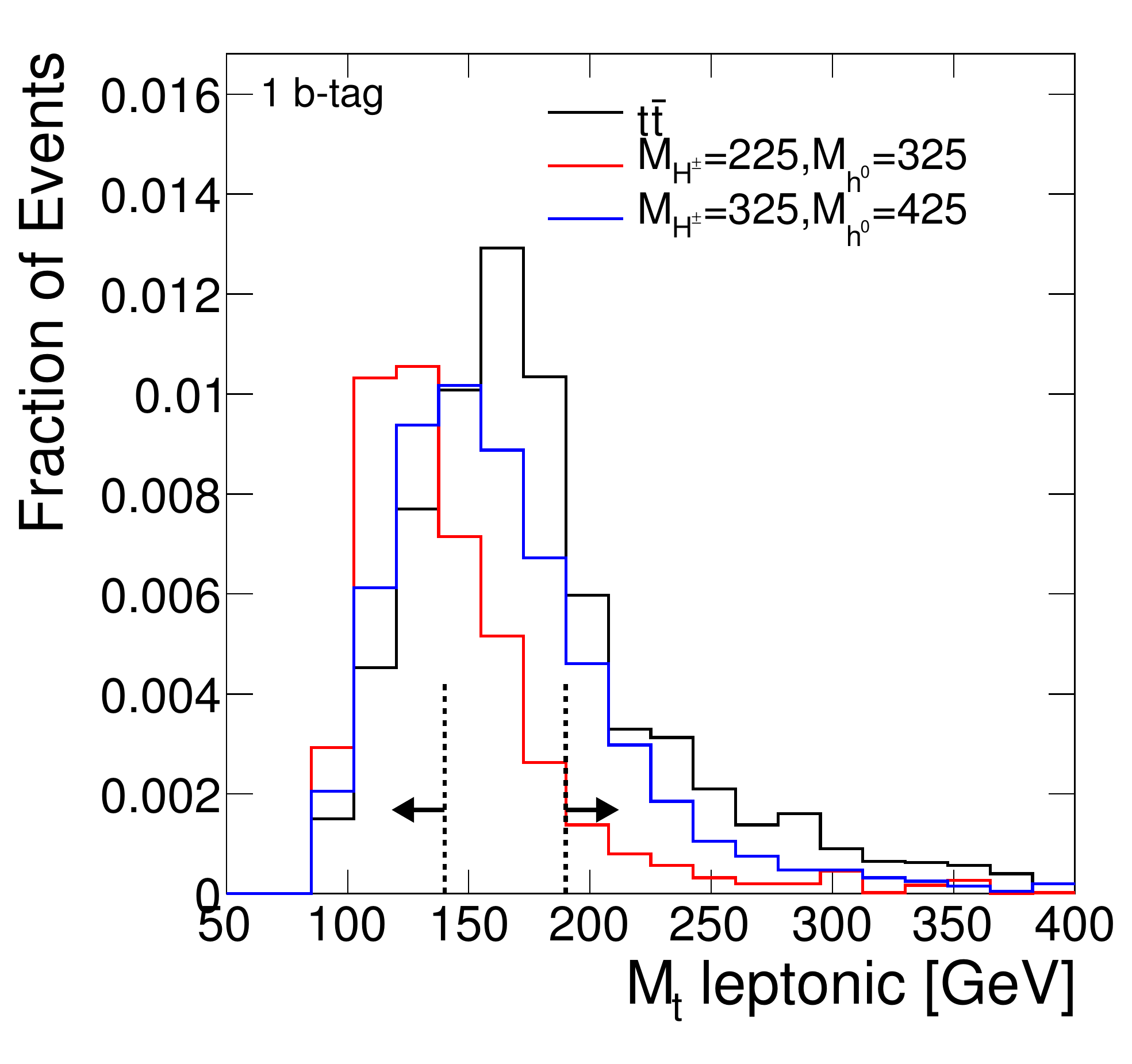}
\includegraphics[width=0.48\linewidth]{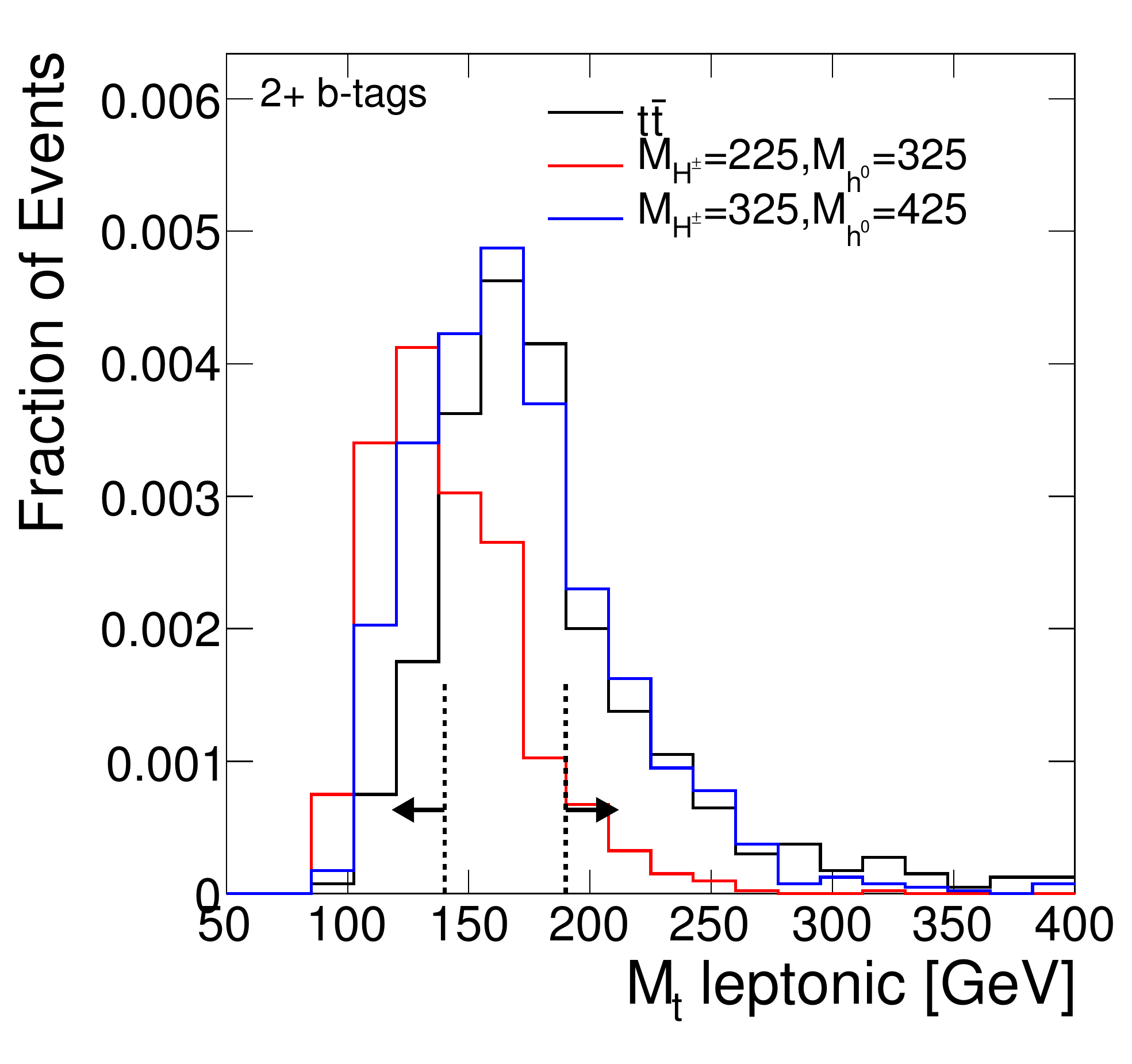}\\
\includegraphics[width=0.48\linewidth]{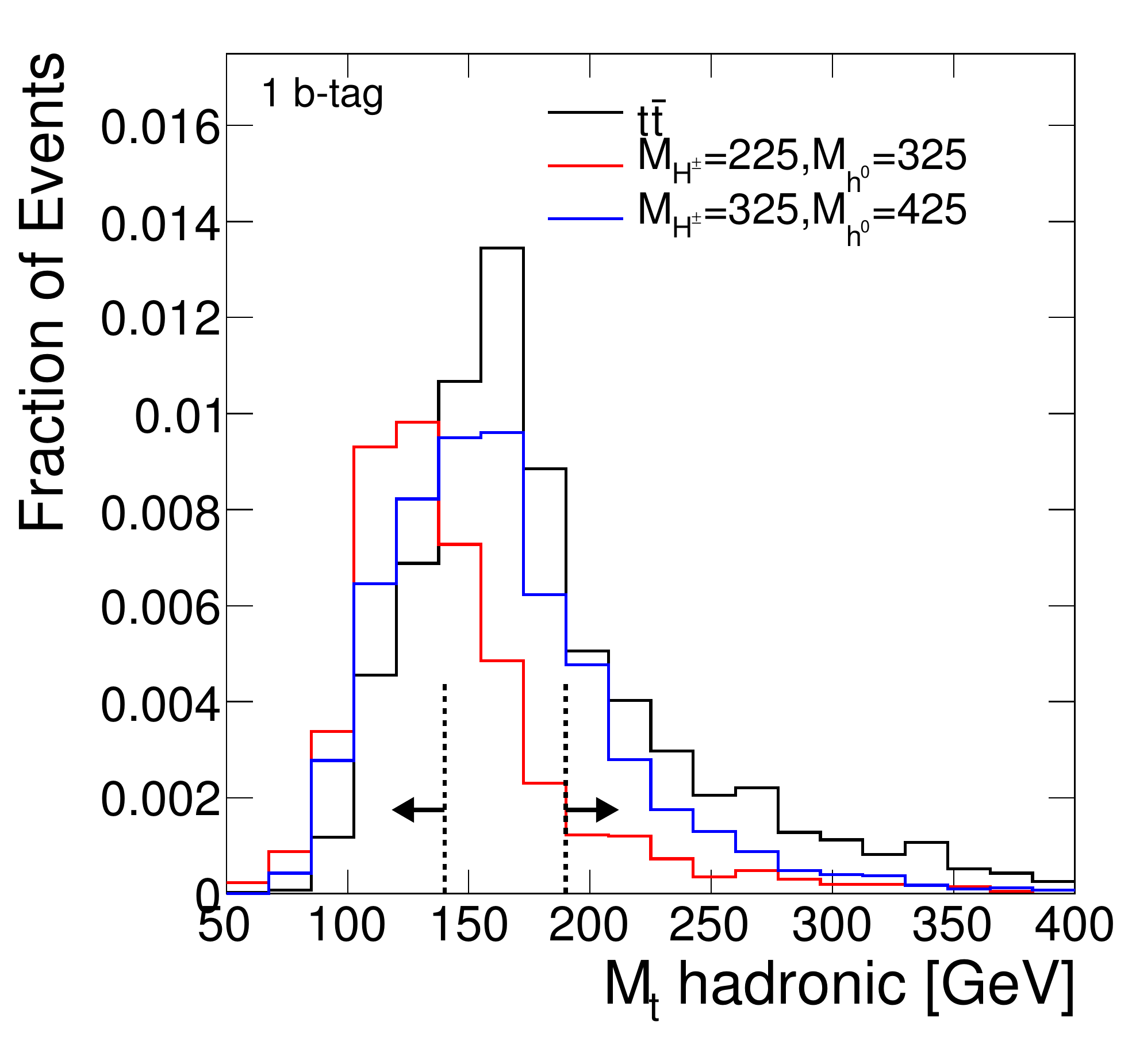}
\includegraphics[width=0.48\linewidth]{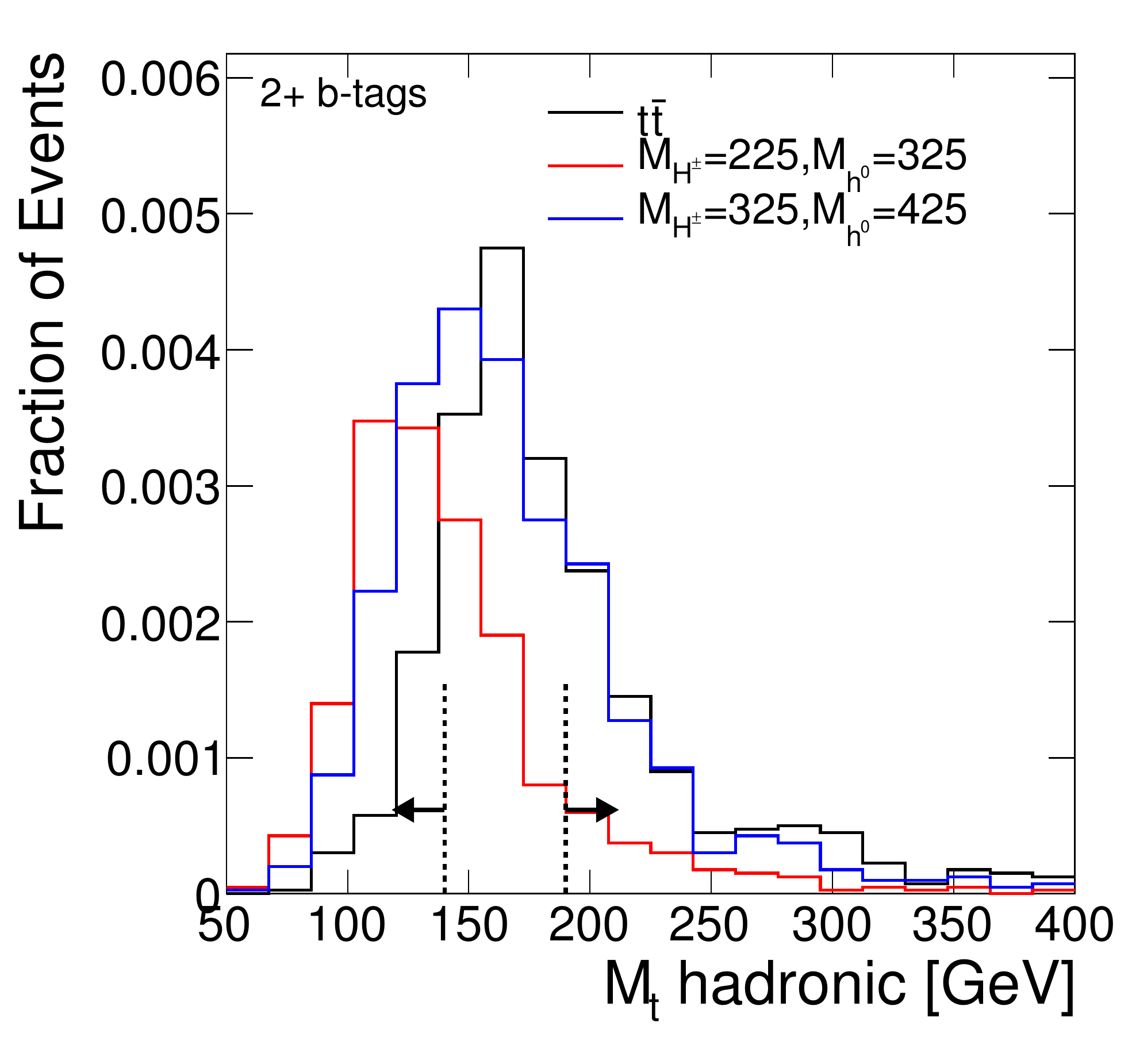}\\
\includegraphics[width=0.48\linewidth]{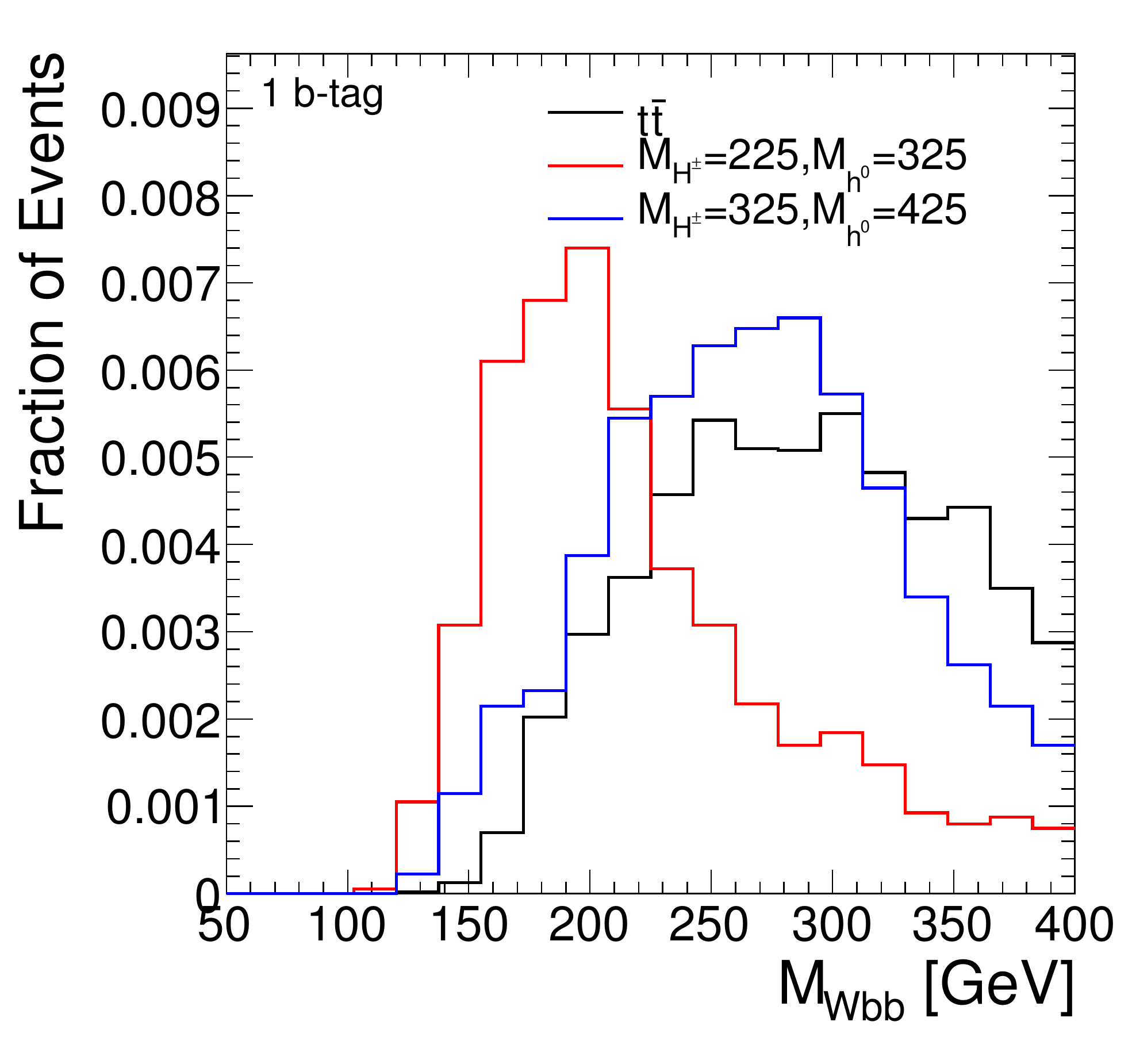}
\includegraphics[width=0.48\linewidth]{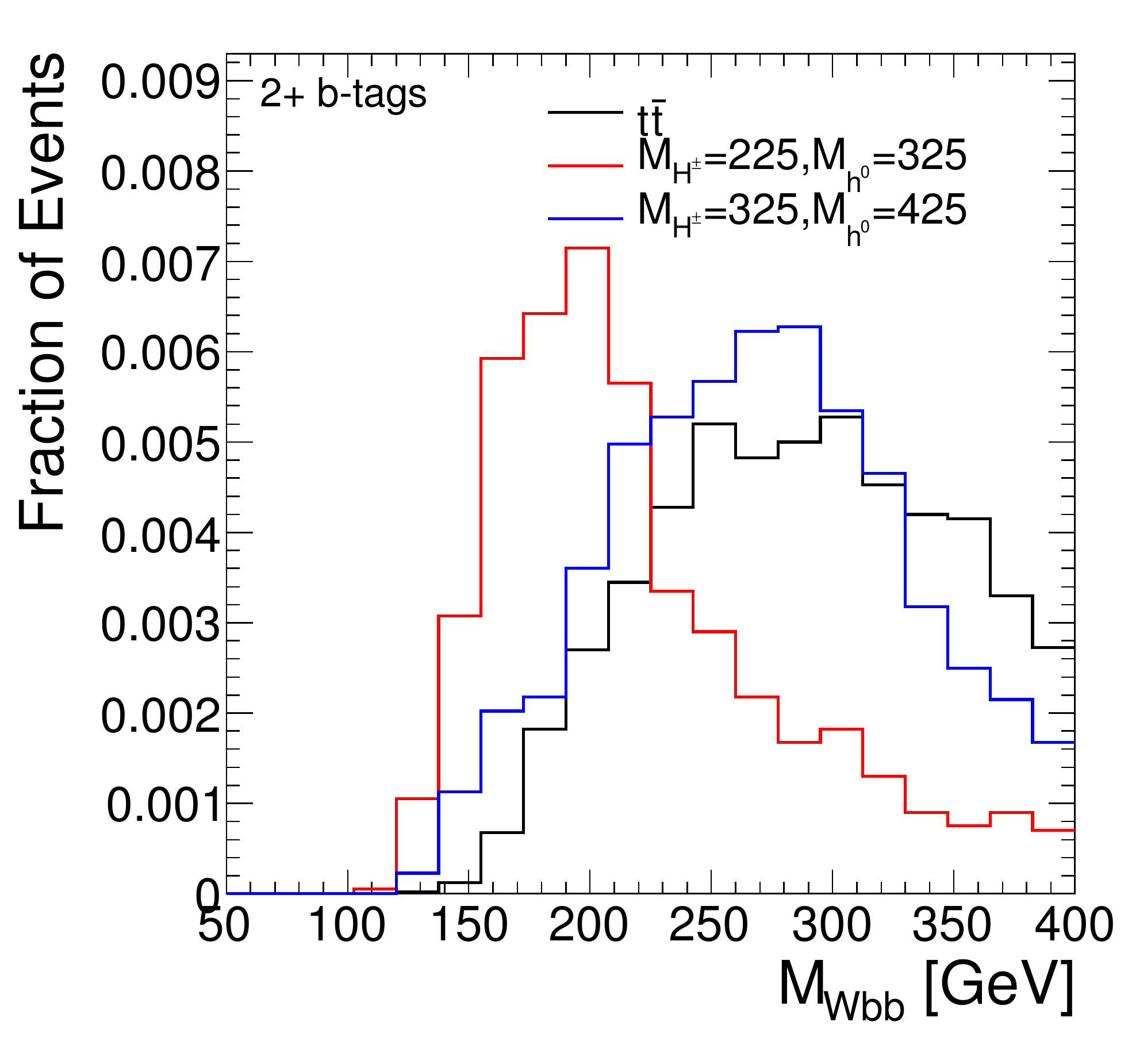}
\end{center}
\caption{ Expected kinematic features of $H^{\pm}\rightarrow W^\pm b\bar{b}$ resonance signal and background events at the
  LHC.  Shown are leptonic top mass (top), hadronic top mass
  (center) and $Wb\bar{b}$ invariant mass (bottom).  Events are categorized
  by the number of $b$-tags seen: left is exactly one tag, right
  is at least two tags.  The top-quark pair background in the $Wb\bar{b}$
  mass distribution is suppressed by a top-quark veto, shown in the
  $M_t$ distributions.}
\label{fig:wbb_kin_lhc}
\end{figure}

\begin{figure}[t]
\begin{center}
\includegraphics[width=0.48\linewidth]{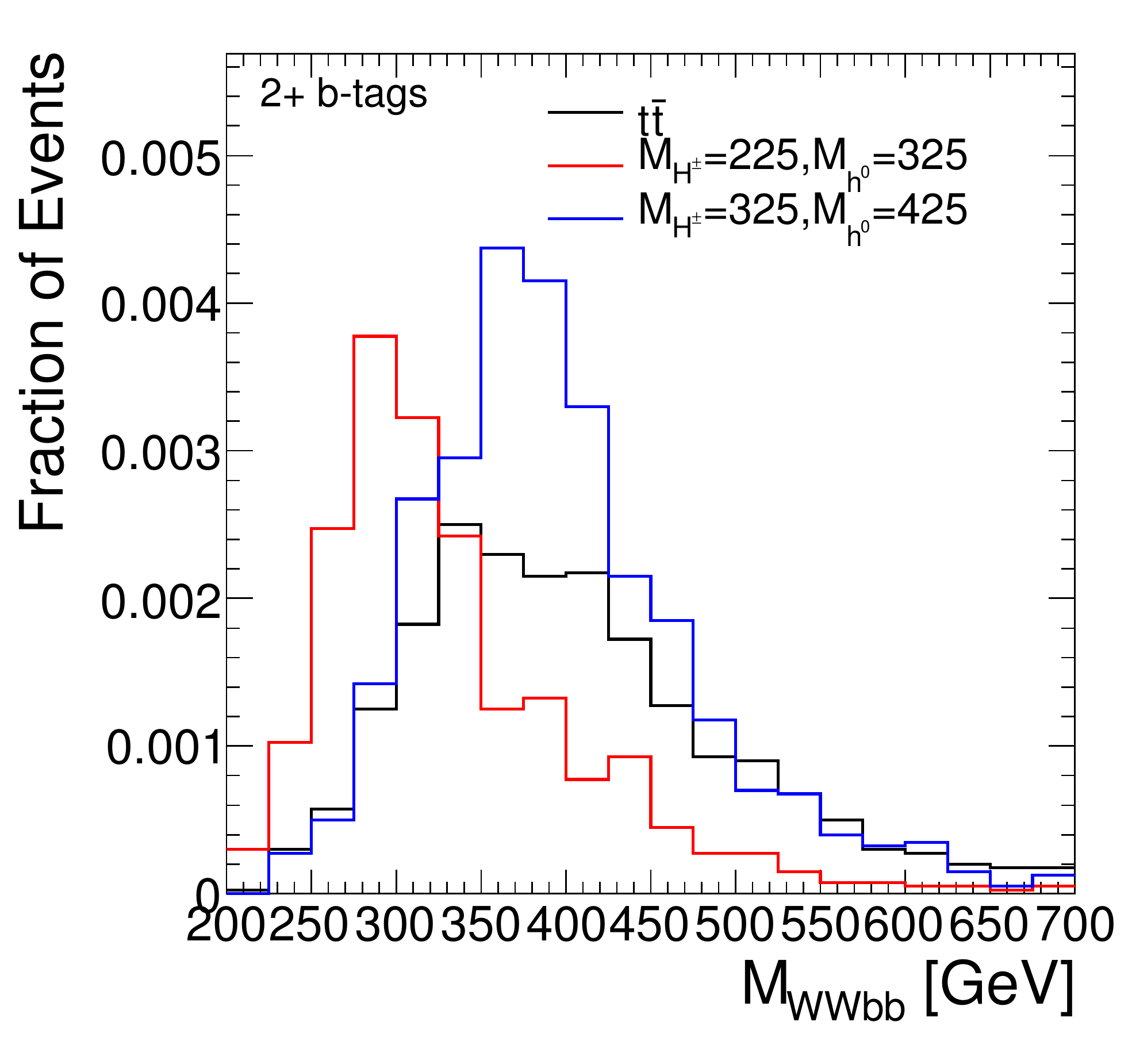}
\includegraphics[width=0.48\linewidth]{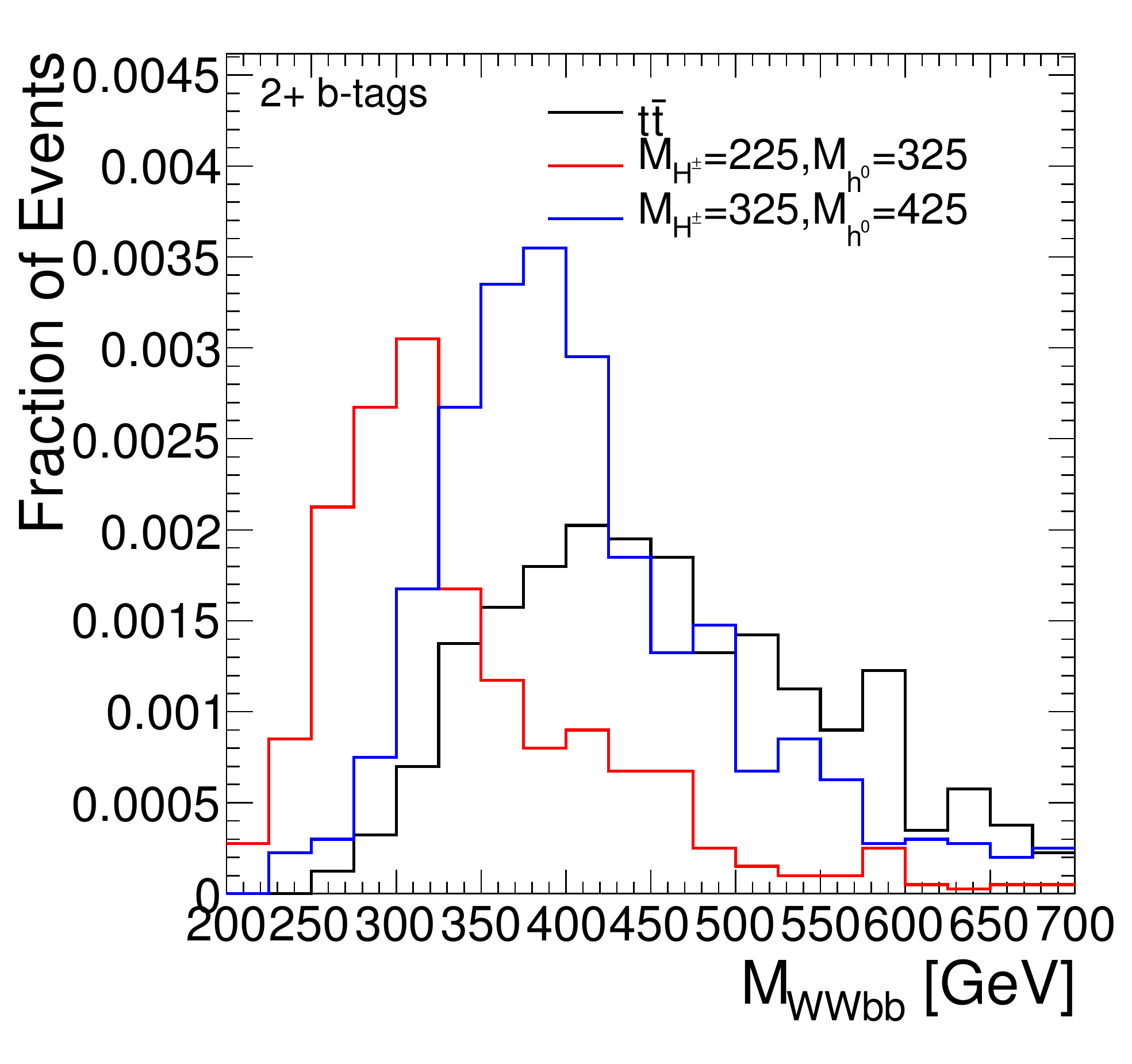}\\
\end{center}
\caption{ Reconstruction of the total invariant mass of the $H^0 \rightarrow H^\pm W^\mp
  \rightarrow Wtb \rightarrow W^+W^-b\bar{b}$ cascade, as
  $m_{WWbb}$. Left is Tevatron, right is LHC.}
\label{fig:wbb_mwwbb}
\end{figure}

\begin{figure}[t]
\begin{center}
\includegraphics[width=0.8\linewidth]{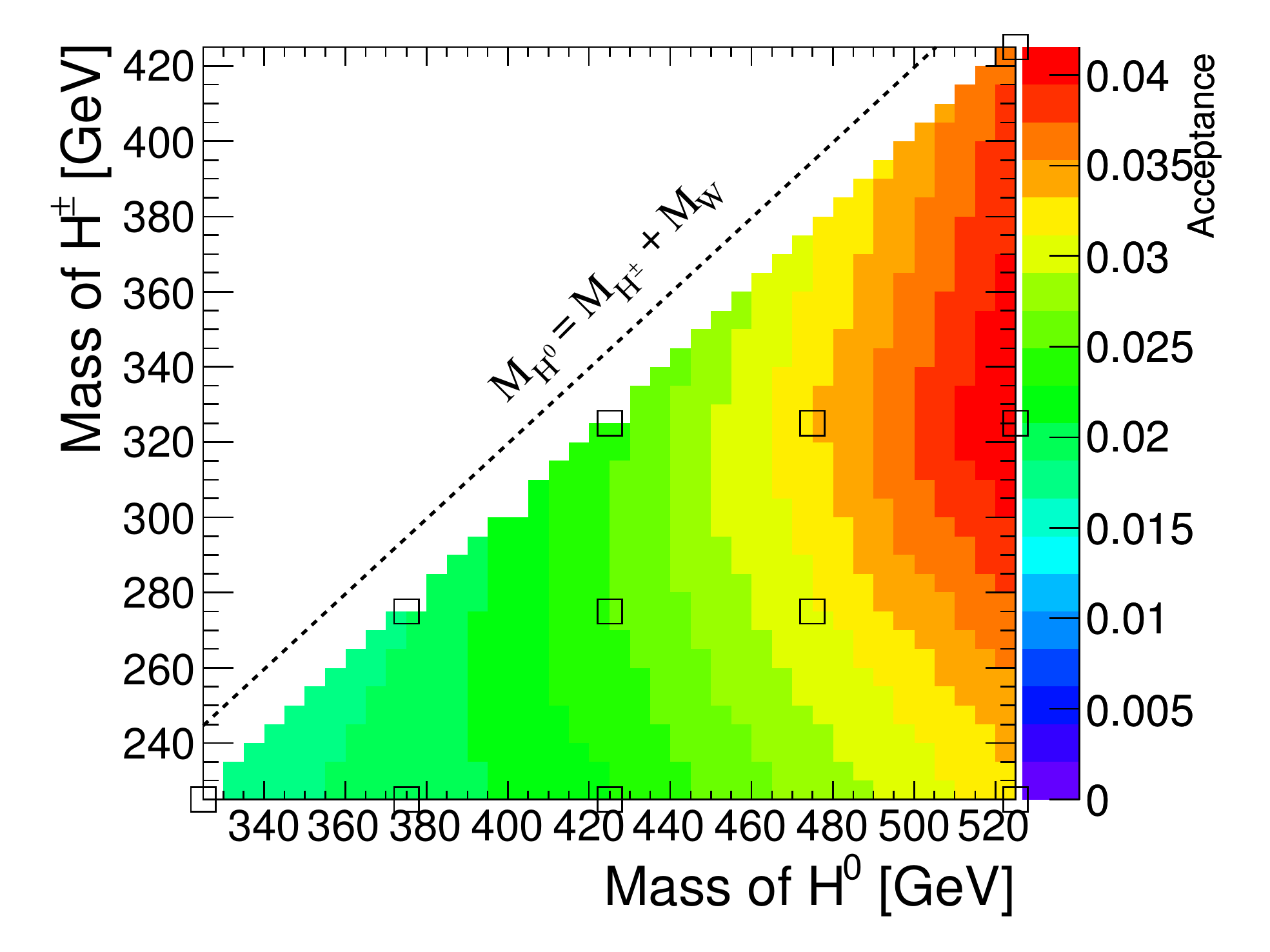}
\includegraphics[width=0.8\linewidth]{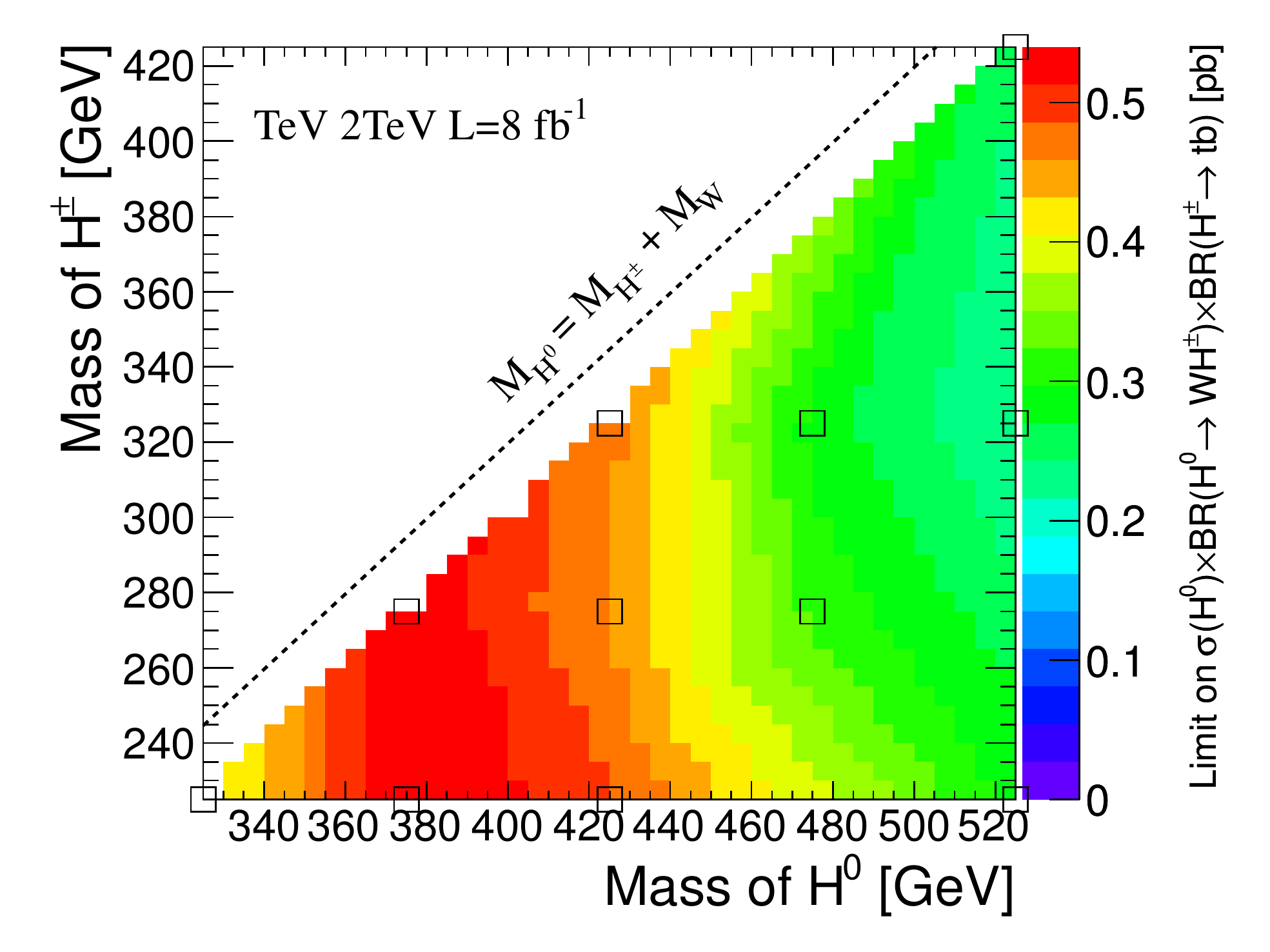}
\end{center}
\caption{ For the Tevatron: top (a) is $H^{\pm}\rightarrow W^\pm b\bar{b}$ signal acceptance after top-quark veto, including the
  branching ratio $W^+W^-\rightarrow \ell\nu qq'$.  Bottom (b) is median
  expected 95\% CL upper limits in the background-only hypothesis.}
\label{fig:wbb_lim}
\end{figure}

\begin{figure}[t]
\begin{center}
\includegraphics[width=0.8\linewidth]{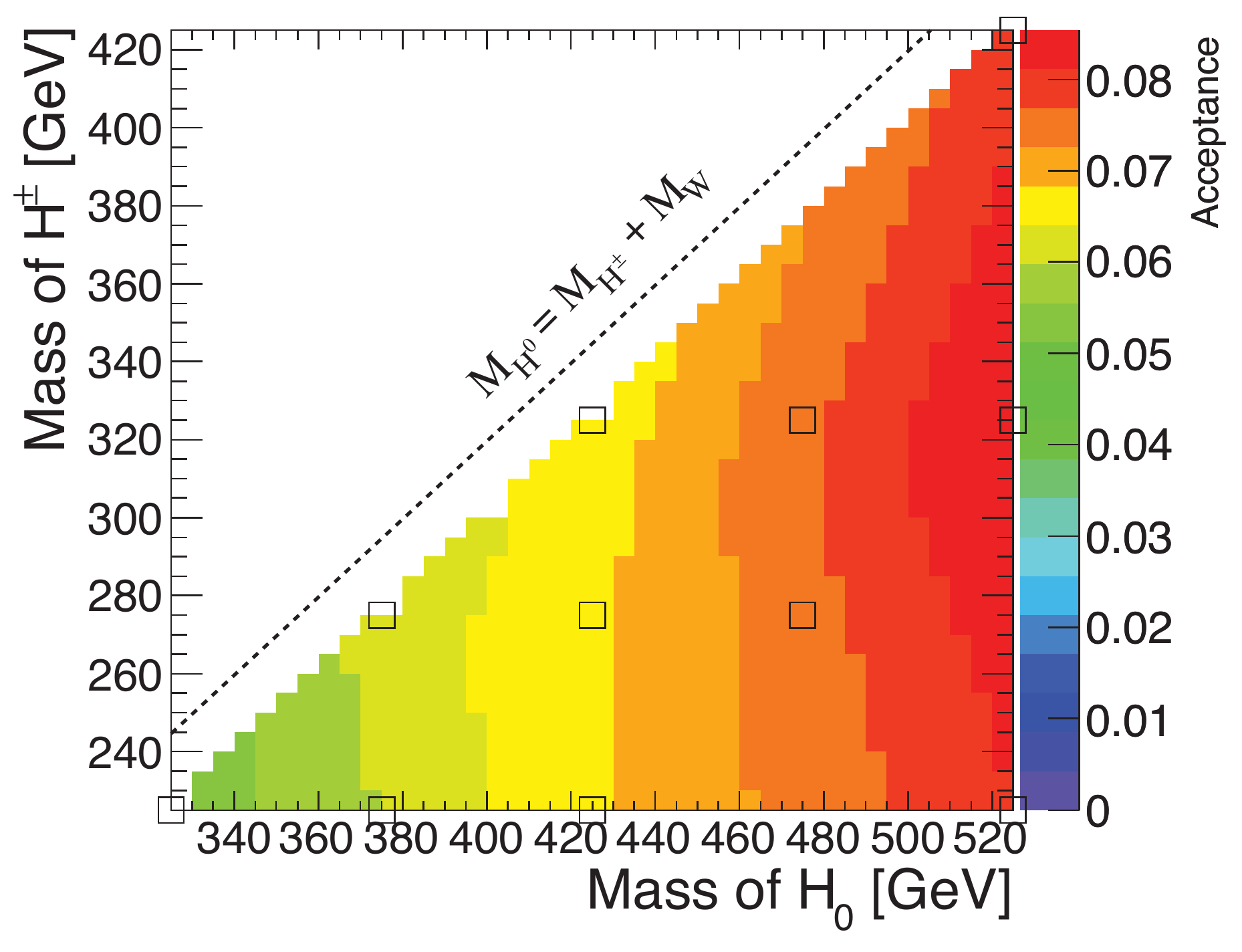}
\includegraphics[width=0.8\linewidth]{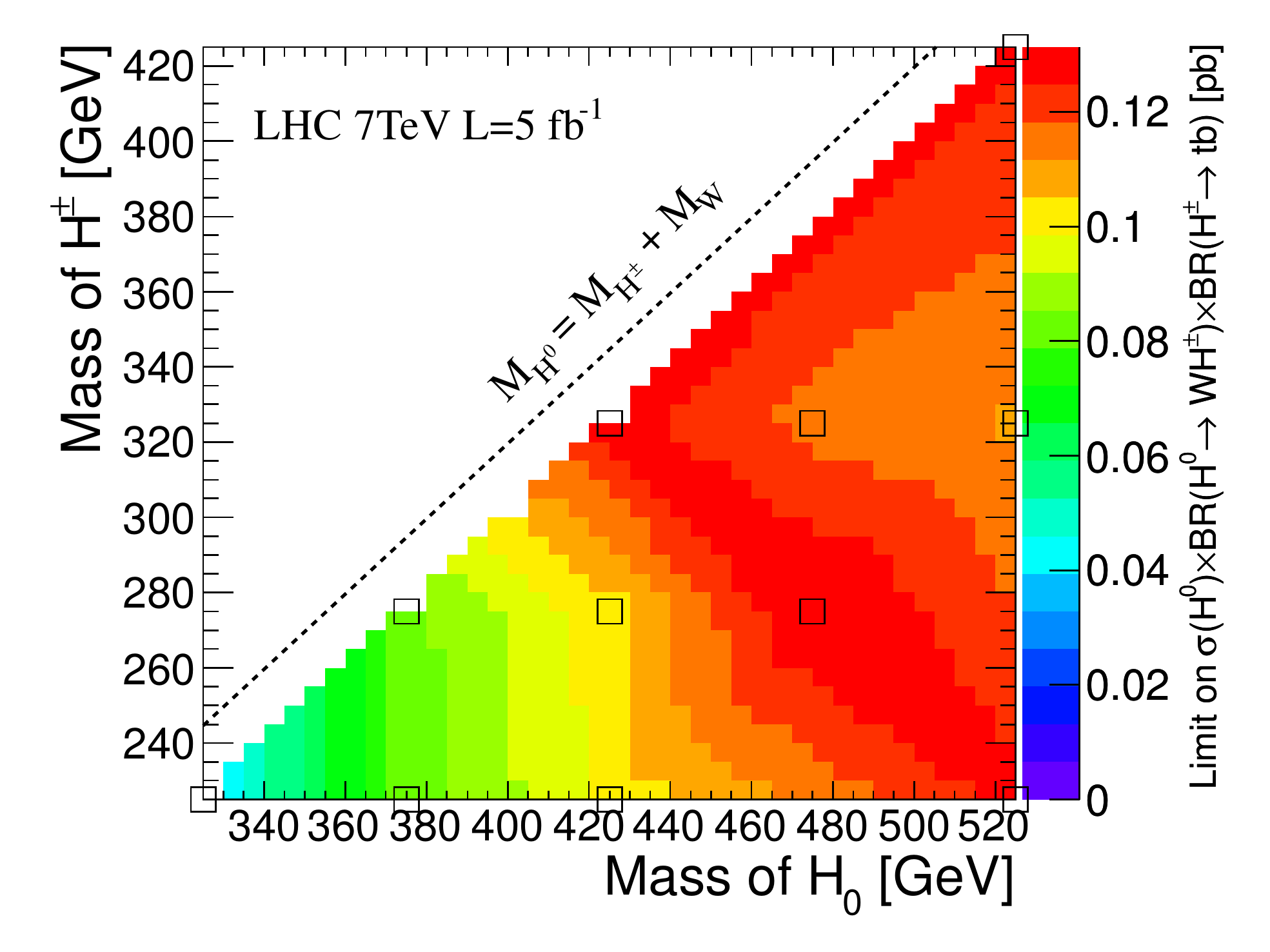}
\end{center}
\caption{ For the LHC: top (a) is $H^{\pm}\rightarrow W^\pm b\bar{b}$ signal acceptance after top-quark veto, including the
  branching ratio $W^+W^-\rightarrow \ell\nu qq'$.  Bottom (b) is median
  expected 95\% CL upper limits in the background-only hypothesis.}
\label{fig:wbb_lim_lhc}
\end{figure}

As in the $b\bar{b}$ resonance case, events are reconstructed according to the $t\bar{t}$ hypothesis, in
order to identify and remove this background.    The $t\bar{t}$
background shows clear peaks in $M_t$ for both leptonic and
hadronic modes, see Fig.~\ref{fig:bb_kin}. The Higgs cascade decay
has exactly one  top-quark decay, leading to reconstructed top-quark masses
that are broader than in the background SM top-quark pair production, but more
difficult to discriminate from $t\bar{t}$ than the $b\bar{b}$ case which has
no top quarks. To reduce the $t\bar{t}$ background, we
veto events if $M_t^{\mathrm{lep}} \in [ M-10,M+10]$ {\bf and}
$M_t^{\mathrm{had}} \in [ M-10,M+10]$, where $M$ is the median
reconstructed $M$ in simulated $t\bar{t}$ events, and the window size
is optimized to maximize expected sensitivity.

The resonance mass is formed by $m_{Wbb}$, choosing the $W$ boson that
gives the largest value of $m_{Wbb}$. It shows a clear peak in
simulated Higgs cascade events, see Fig.~\ref{fig:wbb_kin}. As for
the $b\bar{b}$ resonance, one could likely further improve the
  background rejection by using the $H^0$ resonance in $m_{WWbb}$ as a
  second analysis dimension, see Fig.~\ref{fig:wbb_mwwbb}.

Signal and background yields are calculated as in the $b\bar{b}$ case
described above.  The signal acceptance is calculated using
simulated events, see Figs.~\ref{fig:wbb_lim}a \& \ref{fig:wbb_lim_lhc}a. The median expected upper limit is
extracted in the background-only hypothesis, see Figs.~\ref{fig:wbb_lim}b \& \ref{fig:wbb_lim_lhc}b.

\section{Conclusions}

Extended Higgs sectors can produce Higgs cascade decays leading to a
$W^+ W^- \bar{b}b$ final state that appears in top-quark pair production.
We have shown that the
resonance structure of the Higgs cascade decays can be used to
distinguish these experimentally.
The investigation here uses only the $\bar{b}b$ 
invariant mass, while the reconstructed masses of the $H^0$ and
$H^\pm$ particles in the cascade may also be useful.  

The $M_t$ distributions, which are the primary reconstructed quantities used 
in top-quark mass measurements, appear different for the signals discussed 
than for the $t\bar{t}$ background.  Therefore, should the above signals exist, 
contamination in the $W^+W^-b\bar{b}$ final states would 
lead to measurements of the top-quark mass yielding artificially high or 
low values, depending on the exact mass hierarchy of the $H^0$ and $H^{\pm}$. 
This could generate perceived tension in the standard model precision 
electroweak fits which use the top-quark mass in order to determine the most likely 
Higgs boson mass.   Study of the $t\bar{t}$ final states proposed here could 
disentangle these effects. 

The simple approach presented in this work is sufficient to demonstrate that both the
Tevatron and the LHC have sensitivity to these processes, and we
leave a more sophisticated analysis for the actual experimental searches.

\section{Acknowledgements}

We thank Will Shepherd, David Shih and Scott Thomas for useful conversations.
The authors are
supported by grants from the Department of Energy Office of Science
and by the Alfred P. Sloan Foundation.

\end{document}